\documentclass[preprint,authoryear]{elsarticle}

\usepackage{graphicx}
\usepackage{epsfig}
\usepackage{natbib}

\newcommand{\be}{\begin{equation}}
\newcommand{\ee}{\end{equation}}
\newcommand{\bea}{\begin{eqnarray}}
\newcommand{\eea}{\end{eqnarray}}

\newcommand{\nd}{\noindent}

\begin{document}

\title{Replicator Dynamics of of Cancer Stem Cell; Selection in the Presence of Differentiation and Plasticity}
\author{Kamran Kaveh$^{1}$, Mohammad Kohandel$^{1, 2}$, Siv Sivaloganathan$^{1,2}$}
\address{$^{1}$Department of Applied Mathematics, University of Waterloo, Waterloo, ON, Canada N2L 3G1\\ $^{2}$Center for Mathematical Medicine, Fields Institute for Research in Mathematical Sciences, Toronto, ON, Canada M5T 3J1}

\begin{abstract}
The cancer stem cell hypothesis has evolved into one of the most important paradigms in cancer research. According to cancer stem cell hypothesis, somatic mutations in a subpopulation of cells can transform them into cancer stem cells with the unique potential of tumour initiation. Stem cells have the potential to produce lineages of non-stem cell populations (differentiated cells) via a ubiquitous hierarchal division scheme. Differentiation of a stem cell into (partially) differentiated cells can happen either symmetrically or asymmetrically. The selection dynamics of a mutant cancer stem cell should be investigated in the light of a stem cell proliferation hierarchy and presence of a non-stem cell population. By constructing a three-compartment Moran-type model composed of normal stem cells, mutant (cancer) stem cells and differentiated cells, we derive the replicator dynamics of stem cell frequencies where asymmetric differentiation and differentiated cell death rates are included in the model. We determine how these new factors change the conditions for a successful mutant invasion and discuss the variation on the steady state fraction of the population as different model parameters are changed. By including the phenotypic plasticity/dedifferentiation, in which a progenitor/differentiated cell can transform back into a cancer stem cell, we show that the effective fitness of mutant stem cells is not only determined by their
proliferation and death rates but also according to their dedifferentiation potential. By numerically solving the model we derive the phase diagram of the advantageous and disadvantageous phases of cancer stem cells in the space of proliferation and dedifferentiation potentials. The result shows that at high enough dedifferentiation rates even a previously disadvantageous mutant can take over the population of normal stem cells. This observation has implications in different areas of cancer research including experimental observations that imply metastatic cancer stem cell types might have lower proliferation potential than other stem cell phenotypes while showing much more phenotypic plasticity and can undergo clonal expansion.
\end{abstract}

\maketitle \vspace{10pt}

\section{Introduction}

According to clonal evolution theory, most tumours arise from single cells through multiple genetic alterations accumulated over time. However, the cancer stem cell hypothesis suggests that cancer cells with similar genetic background originate from a transformed cell which can initiate the rest of tumour population \citep{key:CSChyp1}\citep{key:CSChyp2}\citep{key:CSChyp3}\citep{key:morrison}. This subpopulation of tumour initiating cells, known as cancer stem cells, are hypothesized to be a result of somatic mutations of a normal adult stem cell giving it a proliferative advantage and as a result  generating clonal outgrowth in the tissue which leads to the formation of a neoplasm \citep{key:gene1}\citep{key:gene2}\citep{key:vogelstein}. Combination of these mutation/proliferation mechanisms and microenvironmental factors leads to different stages of cancer progression \citep{key:hallmark1}\citep{key:hallmark2}, which results in genetically and phenotypically heterogeneous tumours \citep{key:jdick}.

Stem cells can divide both symmetrically into two daughter stem cells (self-renewal) or two daughter progenitor cells (full differentiation), or asymmetrically into a daughter stem cell and a progenitor cell (partial differentiation) . Progenitor cells then divide hierarchically into a population of fully differentiated functional tissue cells which lack proliferative potential \citep{key:nature-asymmetric}. This polarity in cancer stem cell division is observed in different cancer types. For breast carcinomas, it has been shown that activating the ErbB2 oncogene increases the self-renewal potential of cancer stem cells significantly \citep{key:p53}. Similarly, p53 inactivation leads to not only almost immortal stem cells but also a higher divisional polarity \citep{key:p53}. p53 is also reported to impose an asymmetric proliferation potential on other non-stem cell linages \citep{key:p53-asymmetric}; see also \citep{key:loss-polarity}.

In addition to the tumour initiating potential of cancer stem cells, another distinctive feature of cancer  cells is their high phenotypic plasticity. One aspect of such plasticity is the dedifferentiation potential possessed by stem cell progenitors. During dedifferentiation progenitors (or differentiated cells) can transform back spontaneously into a stem cell thus lending further credence to the vivid concept of cancer stem cells as tumour initiating cell \citep{key:weinberg-rev}. Recent in-vitro experiments have demonstrated the dedifferentiation potential of different cancer type cell lines. For breast cancer, purified
populations of non-stem cells, ${\rm CD44^{\textrm{low}}/CD24^{\textrm{hi}}}$ (basal and luminal cell lines), created a population of ${\rm CD44^{\textrm{hi}}/CD24^{\textrm{low}}}$, which is a marker for stemness \citep{key:gupta}. It has been also shown that a population where the majority are non-stem cells, ${\rm CD44^{\textrm{low}}/CD24^{\textrm{hi}}}$, gives rise to a higher mammosphere formation rate which is a measure of stemness \citep{key:weinberg-pnas}. The role of dedifferentiation in intestinal tumorigenesis is investigated in \citep{key:intestinal-dediff}, where it is shown that elevating the levels of the transcription factor ${\rm NF}-\kappa{\rm B}$, which modulate Wnt signaling, induces dedifferentiation in the (non-stem cell) intestinal epithelial cell population and thus can lead to tumourigenesis. In the context of leukaemia, the leukemia-initiating cell marker ${\rm CD34^{+}CD38^{-}}$ has been observed in the fraction of non-leukemia initiating cells \citep{key:CSChyp1}\citep{key:lukemia-dediff}.

It has been suggested that as cancer progresses towards more aggressive metastatic phenotypes, the dedifferentiation potential increases \citep{key:weinberg-rev}. The dedifferentiation of non-stem cells may arise due to (stochastically) genetic or epigenetic mutations, or the epithelial-mesenchymal transition (EMT), a cellular differentiation process wherein epithelial cells adopt mesenchymal features \citep{key:weinberg-rev}. It has been shown that EMT induced cells have a higher dedifferentiation potential while at the same time they display features resembling stem cell \citep{key:zeb1}\citep{key:scheel-emt} \citep{key:mani-main}. Thus, beside the mutation/clonal expansion model of cancer progression which leads to genomic heterogeneity in the tumour population, inclusion of hierarchal stem cell proliferation and the dedifferentiation potential of cells leads to more phenotypic heterogeneity inside a tumour. More importantly, cell plasticity shadows the concept of stem cells, in the sense that we cannot compare the two populations of normal and cancer stem cells competing via their corresponding proliferation strengths, but rather the population of non-stem cells has to be included in the picture of the selection process and Darwinian evolution of the tumour.

Mathematical models of selection processes so far have treated the selection mechanism among cancer stem cell populations and normal population assuming higher division rate for mutants due to activation and inactivations of oncogenes/tumour suppressor genes that regulate the growth factor signalling pathways inside the cell. These models were able to successfully describe the selection process occurring prior to each new clonal expansion (due to a new mutation). The dynamics of tumour suppressor gene inactivation in particular has been investigated in the literature in detail \citep{key:komarova-main}\citep{key:nowak-tsg}\citep{key:nowak-nature}; see \citep{key:michor-rev} for a thorough review of evolutionary modelling in cancer. However, there has not been much work with regard to stem cell hierarchy proliferation potential and its effect on selection dynamics. Some recent works have
focused on the asymmetric nature of stem cell division. \citep{key:michor-asymmetric} studied the time to fixation of mutant stem cell selection using a simplified birth-death model of two stem cell population which divides asymmetrically, ignoring the population progenitors and differentiated cells (perhaps for simplicity). A recent computational study by Sprouffske et al \citep{key:sprouffske}, has investigated the effect of an asymmetric division scheme for stem cells by simulating stem cells with random fitness and have discussed Darwinian selection and the existence of disadvantageous subpopulations in the formation of neoplasms. Shahriyari and Komarova \citep{key:key:komarova-asym} has also addressed the evolutionary advantage/disadvantages of symmetric versus asymmetric differentiation by constructing a Moran-type process for one and two-hit mutation models and have analyzed the effect of differentiated cell compartment in the effectiveness of the mutations among stem cell compartment. More recently, Jilkine and Gutenkunst \citep{key:key:jilkines} considered a stochastic model for differentiation and dedifferentiation and investigated time to mutation acquisition in the presence of dedifferentiation mechanism for progenitors of stem cells. They also discussed how asymmetric versus symmetric differentiation can affect the efficiency of dedifferentiation process. The cancer stem cell hypothesis has also been applied in the context of drug resistance as an evolutionary process by Leder et al \citep{key:leder}.

The present work aims to provide a general framework to study the selection dynamics of cancer versus normal stem cells by including asymmetric differentiation, in addition to stem cell self-renewal. This introduces a more challenging mathematical model which now contains two competing stem cell populations and a third differentiated cell compartment, which now both stem cells populations are competing with. By introducing a Moran-type stochastic model for this three-compartment model,  we derive replicator-type dynamics for the three populations of cancer stem cells, normal stem cells and differentiated cells as a function of time. We show that the condition for successful invasion by the cancer stem cells not only depends on their higher division rate or lower death rate, but also on differentiation rate or polarity of their asymmetric division. In constructing the birth-death model we assume three independent parameters for the death rates of mutant stem cells, normal stem cells and the differentiated cell population.
An important feature of our model is that dedifferentiation events can be naturally added to the model. We discuss dedifferentiation (assumed only for cancer stem cells) in detail and show that the proliferation advantage is not only a function of relative fitness of two stem cells and their corresponding differentiation rates but also depends on the strength of plasticity and on the population of non-stem cells. We plot a phase diagram between advantageous and disadvantageous regimes in the space of fitness, plasticity and, differentiation probabilities. We show that assuming finite dedifferentiation rates (consistent with numerical estimate from experiments) a seemingly disadvantageous mutant can successfully initiate clonal expansion into a neoplasm.

The paper is organized as follows: In Sec. 2 we formulate a Moran-type model of stem cell differentiation and dedifferentiation and report the replicator dynamics in the presence of differentiation and dedifferentiation potentials. We also discuss the analytical result of the fixation time for this model in the absence of dedifferentiation. In Sec. 3 we discuss numerical solutions of the replicator dynamics and investigate the population dynamics and average time to fixation of a mutant stem cell as one varies division rates of the stem cell population, relative death rates of stem cells and differentiated cells and differentiation probabilities. Similarly, we look at time to fixation and whether a mutant is advantageous or disadvantageous by varying both relative division rates and dedifferentiation potential. In Sec. 4 we discuss the implications for cancer therapeutics and also possible future directions of investigation.

\section{Replicator Dynamics of Differentiation and Dedifferentiation}

\begin{figure}[!h]
\begin{center}
\epsfig{figure=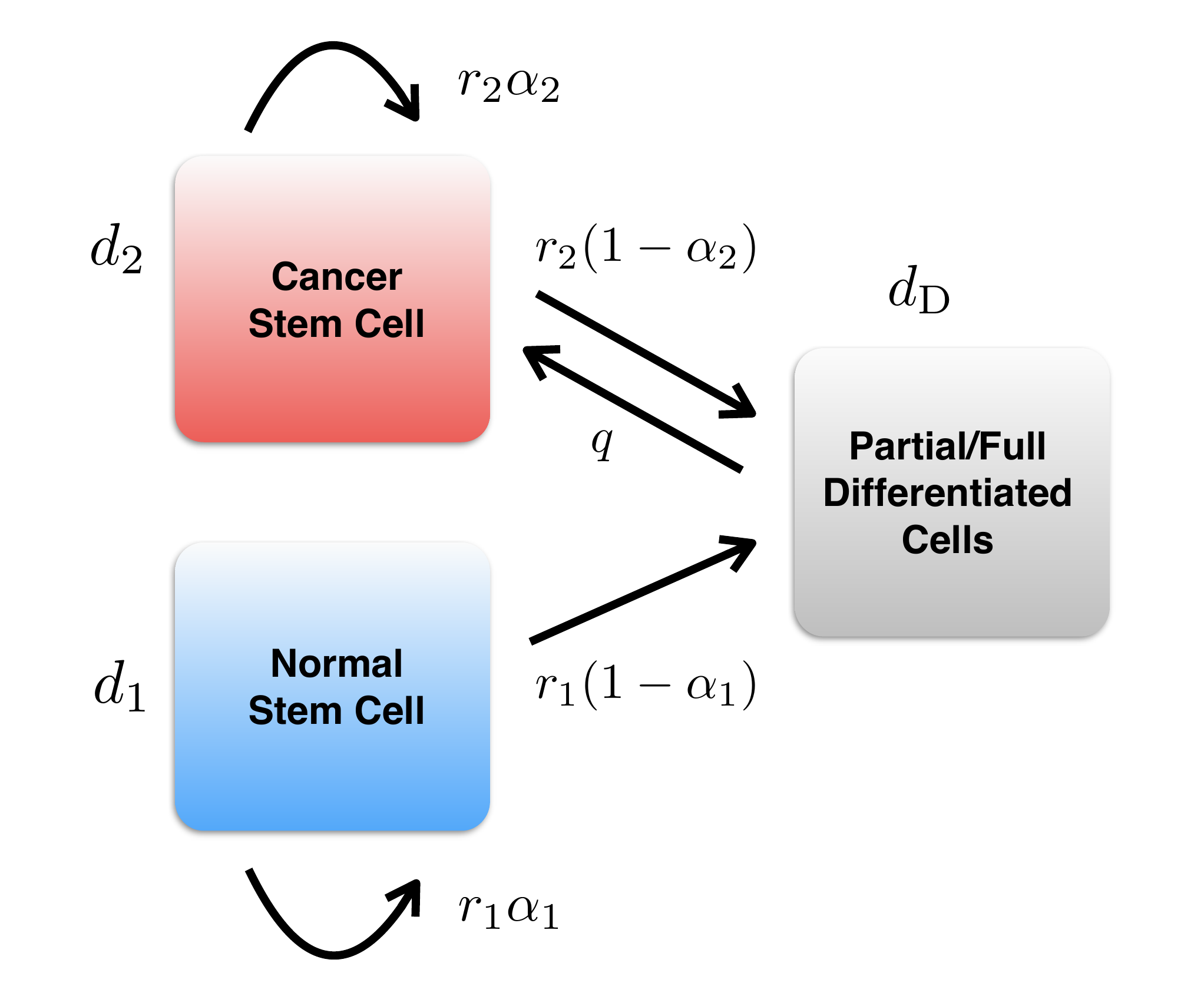, height=240pt,
width=270pt,angle=0}
\end{center}
\caption{The two competing stem cells (normal =blue, mutant = red) can self-renew or differentiate with rates $r_{1,2}\alpha_{1,2}$ and $r_{1,2}(1-\alpha_{1,2})$, and die with rates $d_{1,2}$. The rest of the population of progeny and differentiated cells are indicated as green which can either die, $d_{\rm D}$ or dedifferentiate into the cancer stem cell compartment ($q$).}
\label{three-compartment}
\end{figure}

\begin{figure}[!h]
\begin{center}
\epsfig{figure=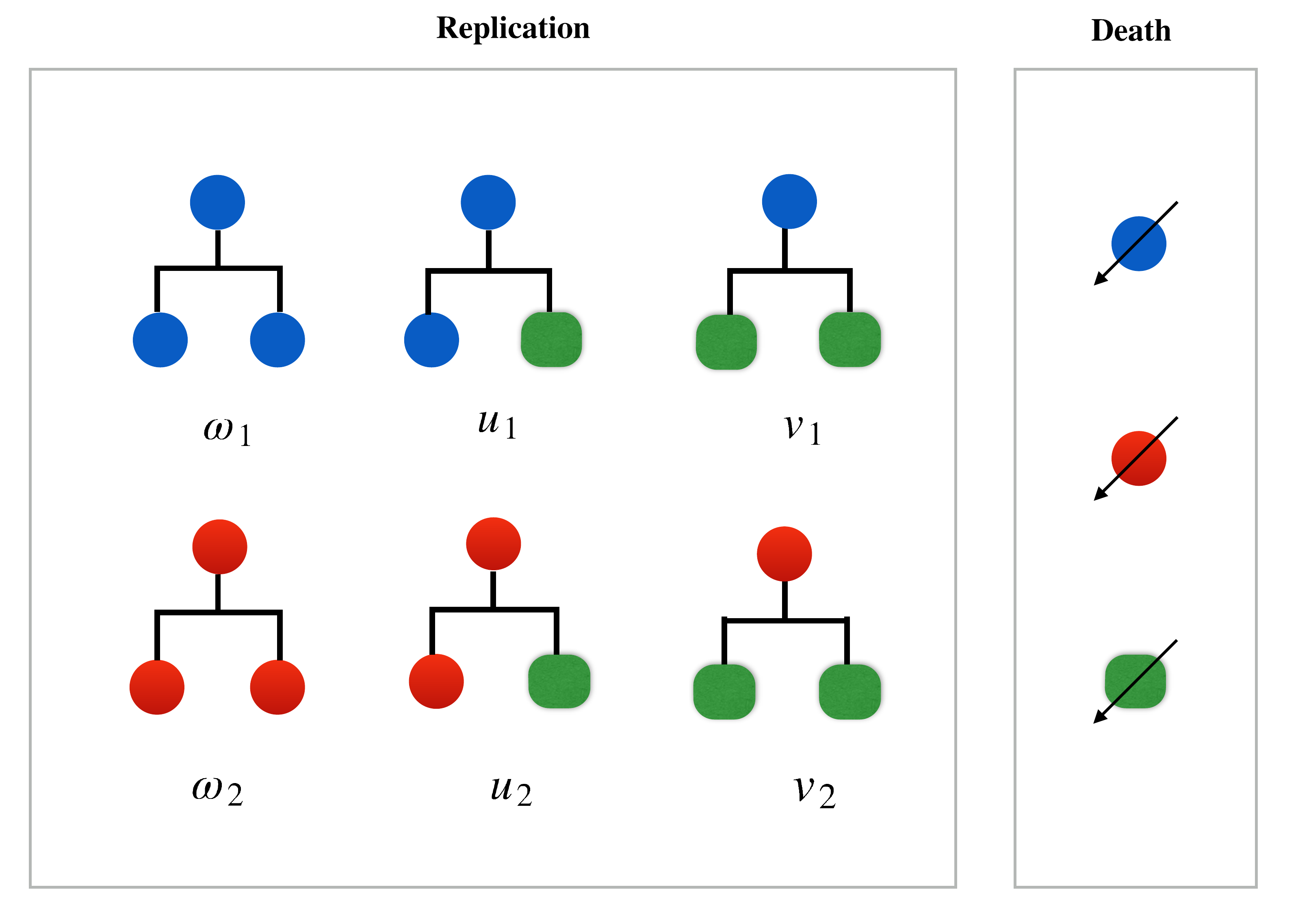, height=240pt,
width=340pt,angle=0}
\epsfig{figure=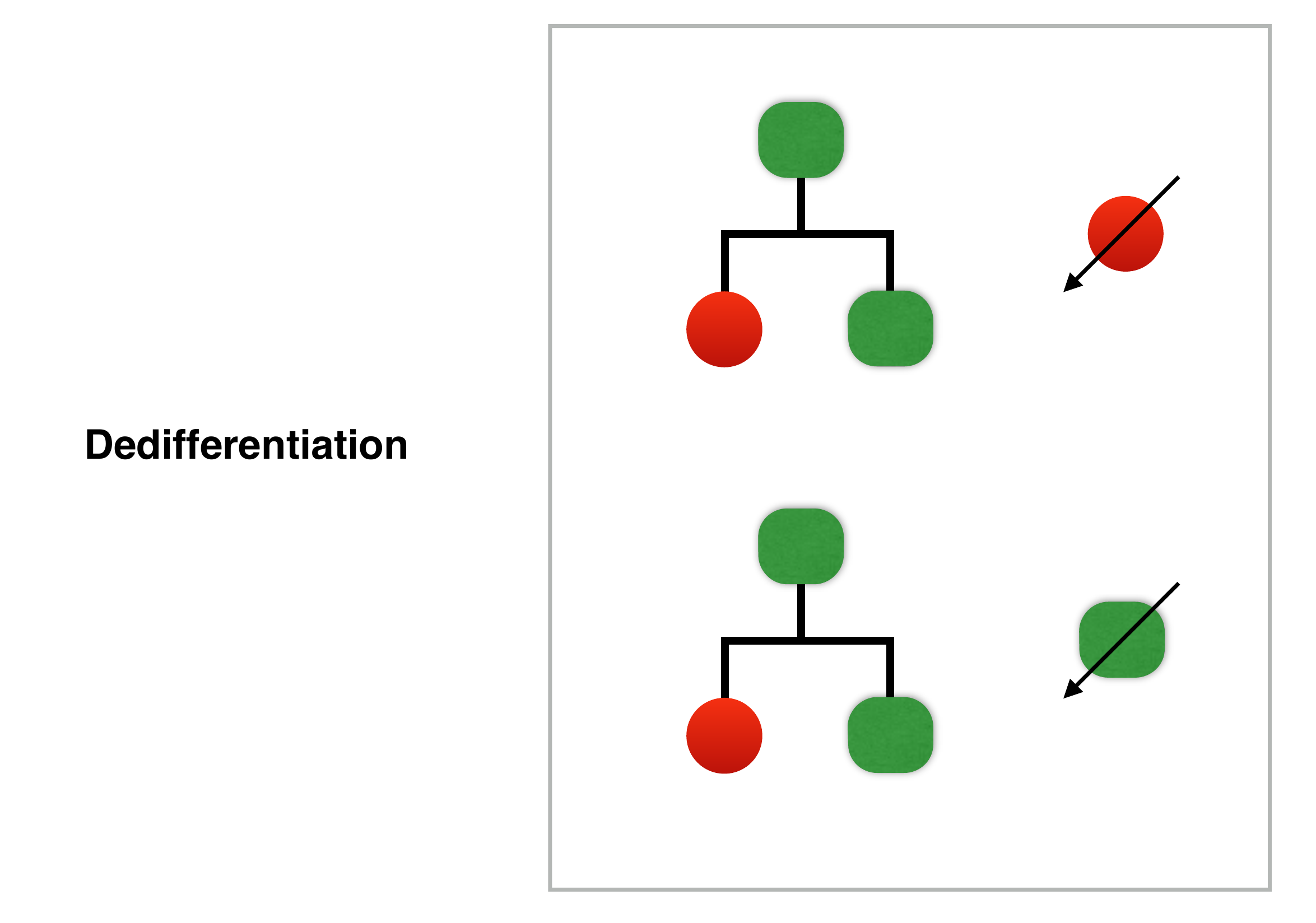, height=180pt,
width=240pt,angle=0}
\end{center}
\caption{The replication-death Moran-type model. The two competing stem cell populations are indicated by blue  and red (normal =blue, mutant = red). The rest of the population of progeny and differentiated cells are indicated as green. In each time step one replication and one death elementary process occurs. In the presence of dedifferentiation, two further composite events happen: a differentiated cell divided into a cancer stem cell and a cancer stem cell death occurs.}
\label{asymmetric}
\end{figure}

We consider a model of two stem cell populations and a population of partially/fully differentiated cells (Fig. \ref{three-compartment}). Normal stem cells divide with a rate $r_{1}$ and die with rate $d_{1}$ per generation.
In each division event, a stem cell can divide (1) symmetrically into two daughter stem cells with probability $\omega_{1}$, (2) asymmetrically into one daughter stem cell and one progenitor with probability $u_{1}$ and (3) fully differentiate into two daughter progenitor cells with probability $v_{1}$ (Fig. \ref{asymmetric}). The probabilities $\omega_{1}, u_{1}$ and $v_{1}$ add up to unity. Similarly, mutant stem cell proliferation and death rates are denoted by $r_{2}$ and $d_{2}$. Correspondingly, self-renewal and differentiation probabilities are $\omega_{2}, u_{2}$ and $v_{2}$. In the hierarchal stem cell proliferation scheme, the number of transient progenitors are
finite thus by including all differentiated cell population into one compartment, we assume that the only possibility for this population to increase is the case in which a stem cell differentiates symmetrically or asymmetrically. Thus we assume that the differentiated compartment lacks proliferation potential of its own. The death rate of differentiated cells are commonly higher than normal and cancer stem cells and is denoted as $d_{\rm D}$.

\begin{figure}[!h]
\begin{center}
\epsfig{figure=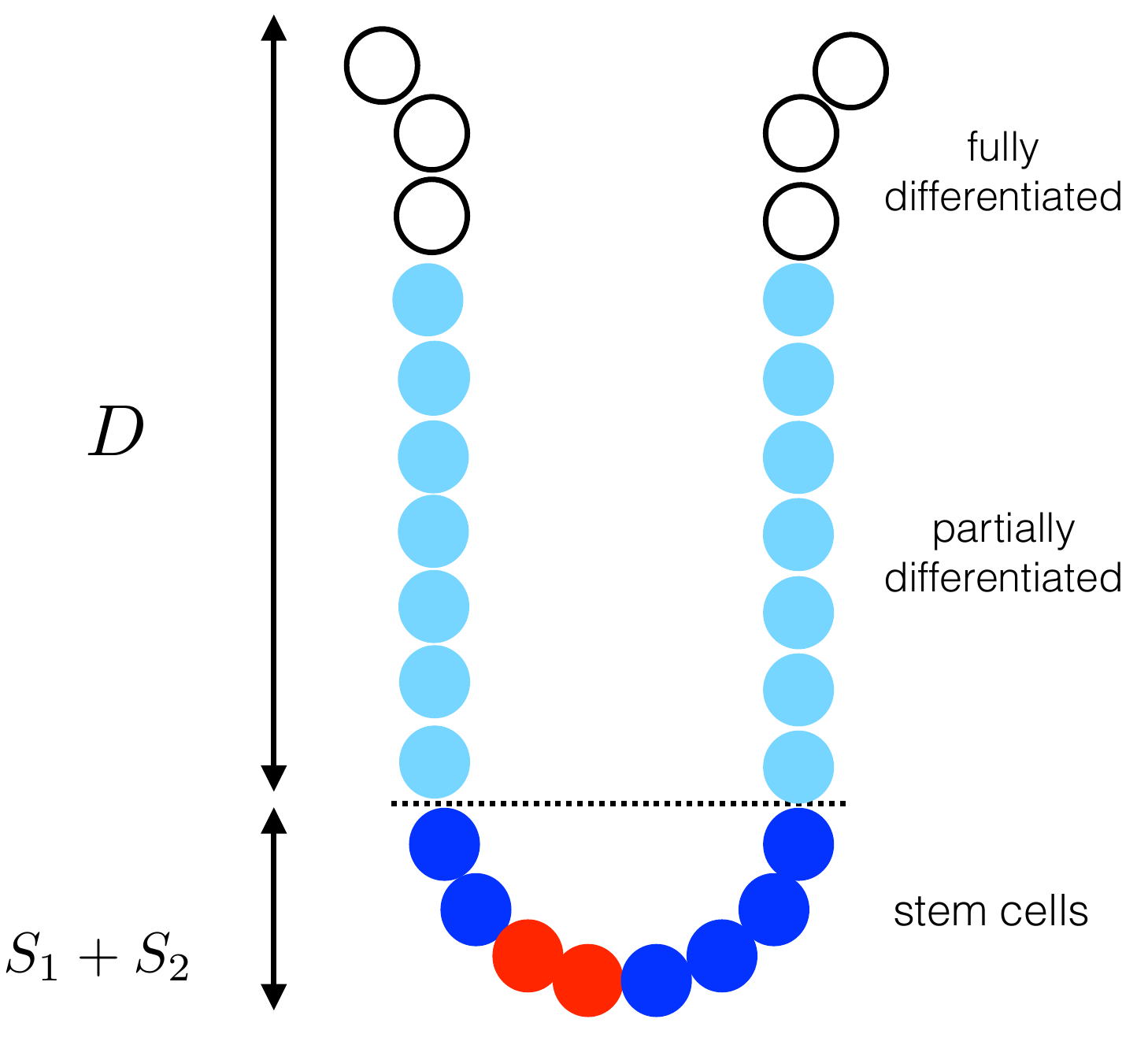, height=150pt,
width=150pt,angle=0}
\end{center}
\caption{Schematics of three-compartment model of cancer/normal stem cell and differentiated cells for an intestinal crypt.}
\label{crypt}
\end{figure}

To biologically justify the model described above, consider an intestinal crypt which constitutes part of the colon. The crypt has a U-shaped one-dimensional arrangement of cells where at the bottom, few stem cells proliferate and augment the proliferation of the rest of the cells. On top of the stem cells, there is a range of transient progenitor cells which constitute most of the body of the crypt and end up to a population of fully functional differentiated cells. Due to both geometrical constraints and homeostasis regulation, the total population of the whole crypt - and not only the population of stem cells at the bottom - can be considered constant (see Fig.~\ref{crypt}).

We consider a Moran-type of birth-death process in which, at every time step, one stem cell is chosen to divide at random with probability proportional to its fitness. The cell division includes all the possible self-renewal and differentiation events. Differentiated cells do not contribute to birth events as they lack any proliferation potential. At the same time step, another stem cell or differentiated cell is picked at random to die. This construction keeps the total population of the whole system of stem cells plus differentiated cells constant. We can examine the dynamics of the selection mechanism by analyzing this model. Notice the differences between this model and the conventional Moran model, where there are only two populations of normal and mutant cells and at every time step a birth and death event occurs. In our model, birth events do not necessarily increase the total number of either of the stem cell populations but rather can be a differentiation process in which the non-stem cell population increases in number, while the number of stem cells is reduced or remains constant. Our construction is also different from that of \citep{key:michor-asymmetric} which incorporated differentiation events as an independent death event.  For a conventional Moran model, the population dynamics is described by a replicator ODE for the fraction of mutant stem cells, $x(t)$ \citep{key:moran}\citep{key:komarova-book},

\be
\frac{{\rm d}x(t)}{{\rm d}t} = \frac{(r-1)x(1-x)}{(1 + (r-1)x)}.
\ee
This can straightforwardly be generalized to arbitrary birth and death rates $r$ and $d$,

\be
\frac{{\rm d}x(t)}{{\rm d}t} = \frac{(\tilde{r}-1)x(1-x)}{\displaystyle\left[ (\frac{r_{2}}{r_{1}}-1)x + 1)\right]\left[(1 - \frac{d_{1}}{d_{2}})x +\frac{d_{1}}{d_{2}}\right]},
\ee

\be
\tilde{r} = \frac{r_{2}}{r_{1}}\frac{d_{1}}{d_{2}}.
\label{fitness-moran}
\ee

If $\tilde{r} > 1$ the mutant is advantageous, if $\tilde{r} < 1$ the mutant is disadvantageous. The case $\tilde{r} = 1$ is the
marginal neutral limit where stochastic fluctuation and finite-population effects are important in determining the fate of a single mutant. In the presence of asymmetric differentiation and the third population of (differentiated) cells, we generalize the above simple criteria for a mutant stem cell being advantageous. The population dynamics of the two stem cell populations can be obtained as a non-linear system of coupled ODEs (see Appendix A),

\bea
 \displaystyle
 \frac{{\rm d}x_{1}(t)}{{\rm d}t} = \frac{r_{1}\alpha_{1}d_{\rm D}\cdot x_{1}-x_{1}d_{1}\left( r_{1}x_{1} + r_{2}x_{2}\right) + r_{1}\alpha_{1} x_{1}((d_{1}-d_{\rm D})x_{1} + (d_{2}-d_{\rm D})x_{2})}{(r_{1}x_{1} + r_{2}x_{2})((d_{1}-d_{\rm D})x_{1} + (d_{2}-d_{\rm D})x_{2} + d_{\rm D})},\nonumber\\
 \frac{{\rm d}x_{2}(t)}{{\rm d}t} = \frac{r_{2}\alpha_{2}d_{\rm D}\cdot x_{2}-x_{2}d_{2}\left( r_{1}x_{1} + r_{2}x_{2}\right) + r_{2}\alpha_{2}x_{2}((d_{1}-d_{\rm D})x_{1} + (d_{2}-d_{\rm D})x_{2})}{(r_{1}x_{1} + r_{2}x_{2})((d_{1}-d_{\rm D})x_{1} + (d_{2}-d_{\rm D})x_{2} + d_{\rm D})},\nonumber\\
\label{replicator}
\eea

\nd where $x_{1,2}$ are the frequencies of the normal and cancer stem cells ($x_{1,2} = i_{1,2}/N$ where $i_{1,2}$ are the corresponding stem cell populations), the coefficients $\alpha_{1,2}=\omega_{1,2}-v_{1,2}$ indicate the effective self-renewal rate. Notice that now $x_{1} + x_{2}
+x_{\rm D} = 1$, where $x_{\rm D}$ is the fraction of the differentiated cells.  Equation \ref{replicator} is the replicator dynamics of caner stem cells in the presence of differentiation.

Attractive fixed points of Eq. \ref{replicator} determine the possible steady state fractions of the two stem cell populations. In the absence of any game theoretical
interactions there is no coexistence of the two populations. Fixed point of Eq. \ref{replicator} corresponding to domination of the mutant population is,

\be
x^{*}_{1} = 0,~~~~~~~~x^{*}_{2} = \frac{\alpha_{2}d_{\rm D}}{(1-\alpha_{2})d_{2} +\alpha_{2}d_{\rm D}} ,
\ee

\nd where the value $x^{*}_{2}$ also indicates the final fraction of mutant stem cells while the rest of the fraction is taken by differentiated cells. The effective proliferation rate of each stem cell population is thus determined by $r_{1,2}\alpha_{1,2}=r_{1,2}(\omega_{1,2}-v_{1,2})$. Similarly, the fixed point of normal stem cells lies along the $x_{1}$-axis and is given by,

\be
x^{*}_{1} = \frac{\alpha_{1}d_{\rm D}}{(1-\alpha_{1})d_{1}+\alpha_{1}d_{\rm D}},~~~~~~~~x^{*}_{2} = 0.
\ee

\begin{figure}[!h]
\begin{center}
\epsfig{figure=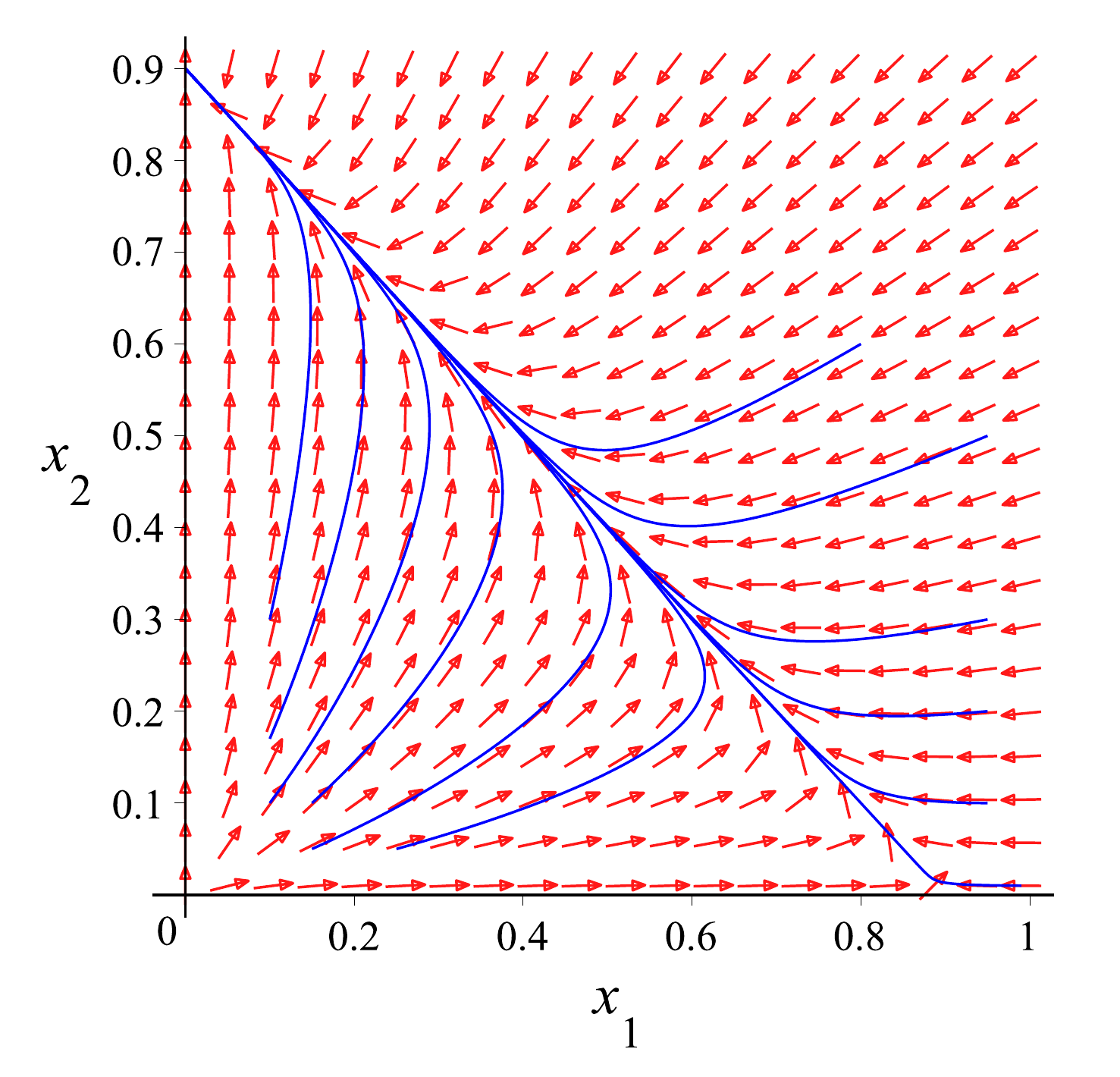, height=170pt,
width=170pt,angle=0}
\epsfig{figure=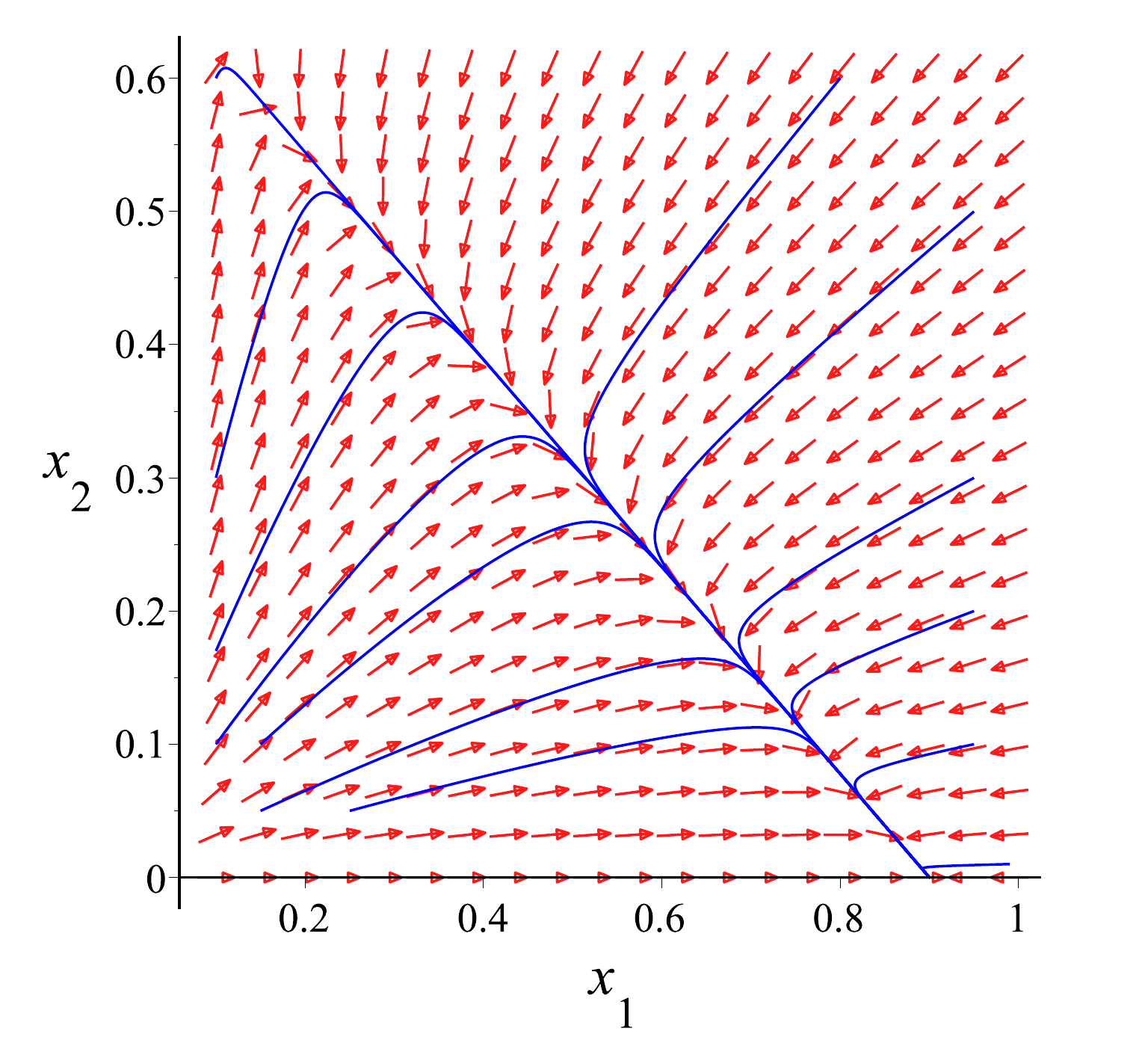, height=170pt,
width=170pt,angle=0}
\end{center}
\caption{The phase portrait of average populations for the stochastic growth equations for (a) $r_{1} =1.0, r_{2} = 1.2, u_{1,2}=0.1, v_{1,2}=0.0, d_{1,2,\rm D}=1.0$ (b) $r_{1}=1.0, r_{2} =1.2, u_{1}=0.1,u_{2}=0.3,v_{1,2}=0.0, d_{1,2,\rm D}=1.0$. Mutant fixed point along vertical axis becomes unstable as the differentiation probability $u_{2}$ increases from 0.1 to 0.3.}
\label{portrait12}
\end{figure}

The phase portrait for Eq. \ref{replicator} is plotted in Fig. \ref{portrait12}. In Fig. \ref{portrait12} one can see that there are two fixed points for this model, $x^{*}_{1,2}$ along each of the two $x_{1}$ and $x_{2}$ axes. It is importance to notice that one of the two fixed points, $x^{*}_{1,2}$ is globally attractive and the other is a saddle-point. The attractive fixed-point determines the advantageous population that is capable of taking over a {\it fraction} of the whole population. As the rest of the population is made of only non-stem cells. A straightforward stability analysis for both fixed points leads to an algebraic condition to determine the advantageous versus non-advantageous phenotypes in the presence of differentiation. One can readily check the following condition for a successful mutant,

\be
\frac{r_{2}(\omega_{2}-v_{2})}{d_{2}} \geq \frac{r_{1}(\omega_{1}-v_{1})}{d_{1}}.
\label{replicator-fitness}
\ee

It turns out that the change in the differentiated cells death rate (which in fact is representative of how fast transient cells are dividing and fully-differentiated cells are dying) does not change the critical value of the effective fitness that determines a neutral system. Variation in the value of $d_{\rm D}$, however, does change the fraction of mutant stem cells in the stationary state and also affects how long it takes to reach fixation (time to fixation).

The denominator of Eq. \ref{replicator} is quite essential in determining the time to fixation for an advantageous mutant. One can derive (see Appendix B) an analytical result for this fixation time in the case of equal death rates $d_{1,2,\rm D} = 1$,
\be
t_{F} = \left(\frac{r\alpha_{2} + \alpha_{1}}{r\alpha_{2}-\alpha_{1}}\right)\ln N - \ln \alpha_{2},
\label{t-fix}
\ee

\nd where $1/N$ is the initial fraction of the mutant stem cells (see Appendix B). As differentiation probabilities increase the second term contribution abruptly increases leading to higher time to fixation. Interestingly $u_{1,2}=v_{1,2}=1/3$ is the limit where time to fixation blows up to infinity.

It is also instructive to follow the trajectory, starting with one mutant stem cell ($x_{2,i} = 1/N$). As indicated by a horizontal trajectory in Fig. \ref{portrait12}, the frequency of mutant cells remain constant indicating that initially differentiation of normal stem cells is a major mechanism until the trajectory gets close to the fixed point of normal cells, i.e. differentiated cells get saturated. The rest of the trajectory is a diagonal line that connects the two fixed points to each other. For the case of identical differentiation probabilities, $\alpha_{1} = \alpha_{2}$ we get $x_{2} = \alpha - x_{1}$, indicating that the selection dynamics, similar to the Moran model, is at work, and the effect of differentiation is a shift in the effective total population ($N \rightarrow N-n_{\rm D}$). In the case of different differentiation rates which is of interest to us, we can see the line of attraction does have a slope and thus differentiation effect acts to change the effective  total population and also to amplify the frequency of the mutant population.

Beginning from a small differentiated cell fraction (see Fig.\ref{portrait12}), we see that the trajectories tends to stay at an almost constant
value for $x_{2}$, while $x_{1}$ is decreasing, i.e the competition is initially between differentiated cells and normal cells. As time passes and the differentiated cell population reaches a saturated value the effective competition is going to be between the two stem cell populations. In three-population models of selection, this sequential behaviour for selection is known as Hill-robertson interference (for example see \citep{key:durrett08}). This is in fact an important observation, as the model predicts that there is a always a (relatively large) pool of differentiated cell population present as the two stem cell populations are competing thus when dedifferentiated events are going to be taken into account (next section).

\subsection{Dedifferentiation and Plasticity}

We can introduce dedifferentiation into this model in a similar fashion as a back-mutation event for differentiated cells. To incorporate mutation events into a Moran process, we can regard it as an asymmetric division event where one phenotype asymmetrically divides and gives rise to one daughter cell of its own type and one
daughter cell of the mutant phenotype (see \citep{key:komarova-main}\citep{key:komarova-book}). In the case of plasticity - as back mutation- we assume that a differentiated cell can asymmetrically divide into a daughter differentiated cell and a daughter mutant stem cell. The rate of division is indicated by $q$. This is accompanied by a death event of either a differentiated cell or a mutant stem cell with the same death rate as model indicates. The slight difference that appears in our construction is that we need a mutation event for differentiated cells to transform back into mutant stem cells and into the normal population. Thus, only the differentiated cell division are accompanied with either mutant or differentiated cell death events. Thus there is a correlation between birth-death dedifferentiation events by construction. As normal stem cells are assumed to be well regulated and follow the stem cell proliferation hierarchy we ignore the possibility of dedifferentiation for the normal population.

To reduce the number of parameters, we assume the death rates of all populations are equal, $d_{1,2,\rm D}=1.0$. Now the population dynamics can be written in terms of the ratio of $r_{2}$ and $r_{1}$ and the dedifferentiation rate, $q$. Derivation of the transition probability for the dedifferentiation events are presented in Appendix A,
\bea
 \displaystyle
 \frac{{\rm d}x_{1}(t)}{{\rm d}t} &=& \frac{r_{1}\alpha_{1}\cdot x_{1}-x_{1}\left( r_{1}x_{1} + r_{2}x_{2}\right)}{((r_{1}-q)x_{1} + (r_{2}-q)x_{2} +q)},\nonumber\\
 \frac{{\rm d}x_{2}(t)}{{\rm d}t} &=& \frac{r_{2}\alpha_{2}\cdot x_{2}-x_{2}\left( r_{1}x_{1} + r_{2}x_{2}\right)+ q(1-x_{1}-x_{2})^{2}}{((r_{1}-q)x_{1} + (r_{2}-q)x_{2}+q)},
\label{plasticity}
\eea

\nd where the term $q (1- x_{1}-x_{2})^{2}$ is added to the dynamics of mutant cells. The dynamics of normal stem cells are untouched other than modification to the total possible events (denominator of equations) which now includes an effective proliferation event of differentiated cells with rate $q$.

In the presence of dedifferentiation, the fate of a new mutant is not only determined by its relative proliferation rate, $r$ but also by
the dedifferentiation rate of non-stem cell population, $q$. Notice that the differentiation dynamics is essential and has to be included, as the differentiation rates essentially determine what fraction of the total population is made up of differentiated cells and thus affect the dedifferentiation rate. For example if differentiation rates are low and we begin with a population of normal stem cells and introduce a disadvantageous mutant to this system, the mutant will become extinct. However, if the differentiation rates are high enough the population of differentiated cells - created via differentiation - can actively contribute to increasing the mutant population.

In the construction of the model we did not make a distinction between differentiated cells with the capability of phenotypic plasticity and normal differentiated cells. Dividing the non-stem cell population into plastic and non-plastic populations, however, will only change the effective value of the
dedifferentiation rate, $q$, and is not crucial for our analysis. For stochastic analysis of the fixation probability, however, such a distinction might be needed.

In the presence of dedifferentiation, the fixed point of mutant cells remains along the $x_{2}$-axis, given by,

\be
x^{*}_{1} = 0,~~~~~~~~x^{*}_{2} = \frac{1}{2}\frac{-\alpha_{2}r_{2}+2q +\sqrt{\alpha_{2}r_{2}(\alpha_{2}r_{2} - 4q)+4qr_{2}}}{q-r_{2}},
\label{fp-dediff}
\ee

\nd while the (previous) fixed point of normal cells now shifts up from the $x_{1}$ axis, having both components nonzero. The components of this fixed point can be described by algebraic expressions in terms of $r_{1,2}, \alpha_{1,2}$ and $q$. In the limit of $\alpha_{1}=1$ and $r_{1}=1$,

\be
x^{*}_{1} = \frac{(\alpha_{2}+q)r^{2}_{2}-2qr_{2}+q-r_{2}}{q(r_{2}-1)^{2}},~~~~~~~~x^{*}_{2} =\frac{-\alpha_{2}r_{2}+1}{q(r_{2}-1)^{2}}.
\label{fp-dediff2}
\ee

Notice that even in the absence of both stem cell populations at $t = 0$ the system will evolve to one of the two fixed points which contain stem cells, due to phenotypic plasticity.

\begin{figure}[!h]
\begin{center}
\epsfig{figure=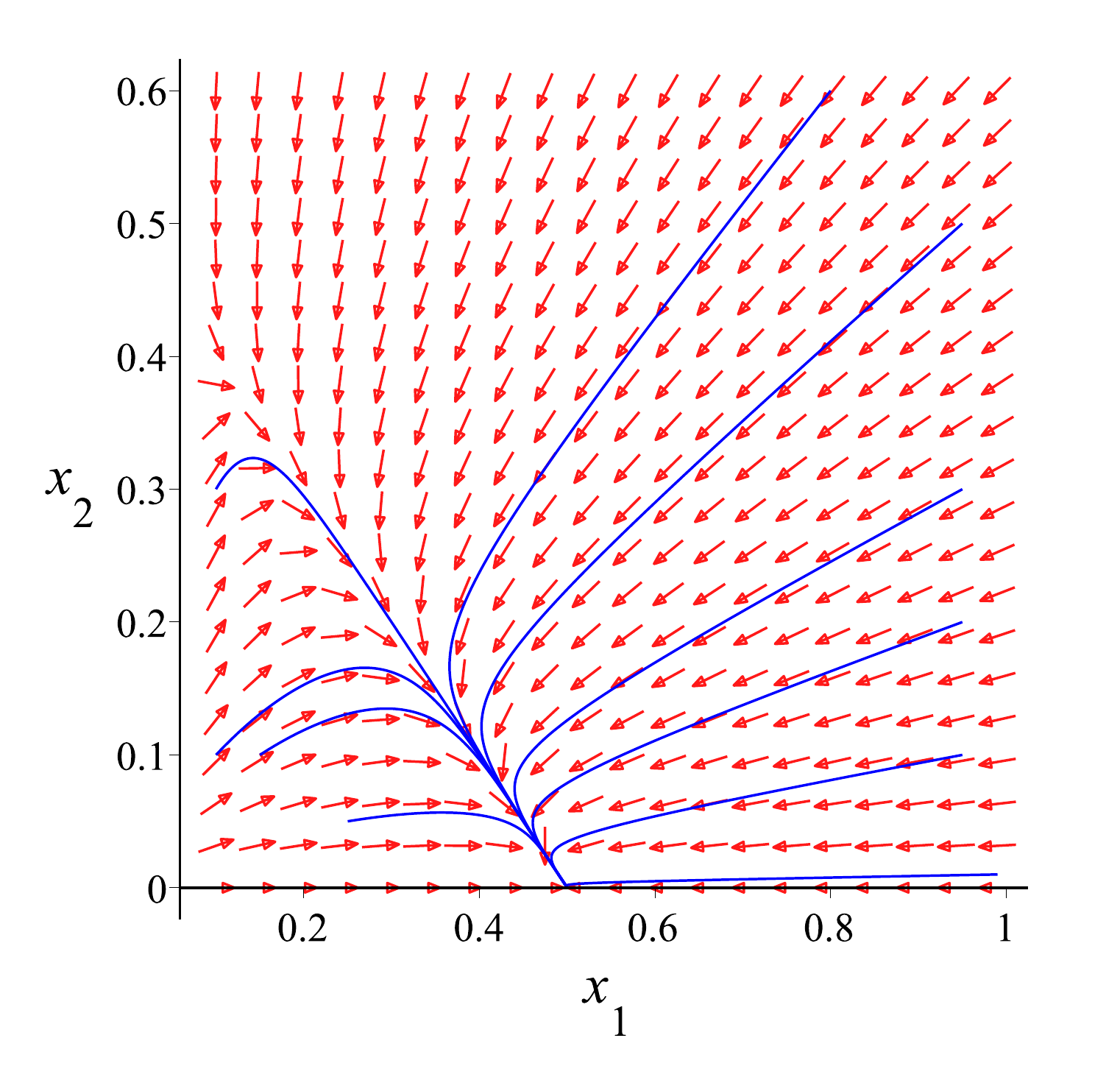, height=170pt,
width=170pt,angle=0}
\epsfig{figure=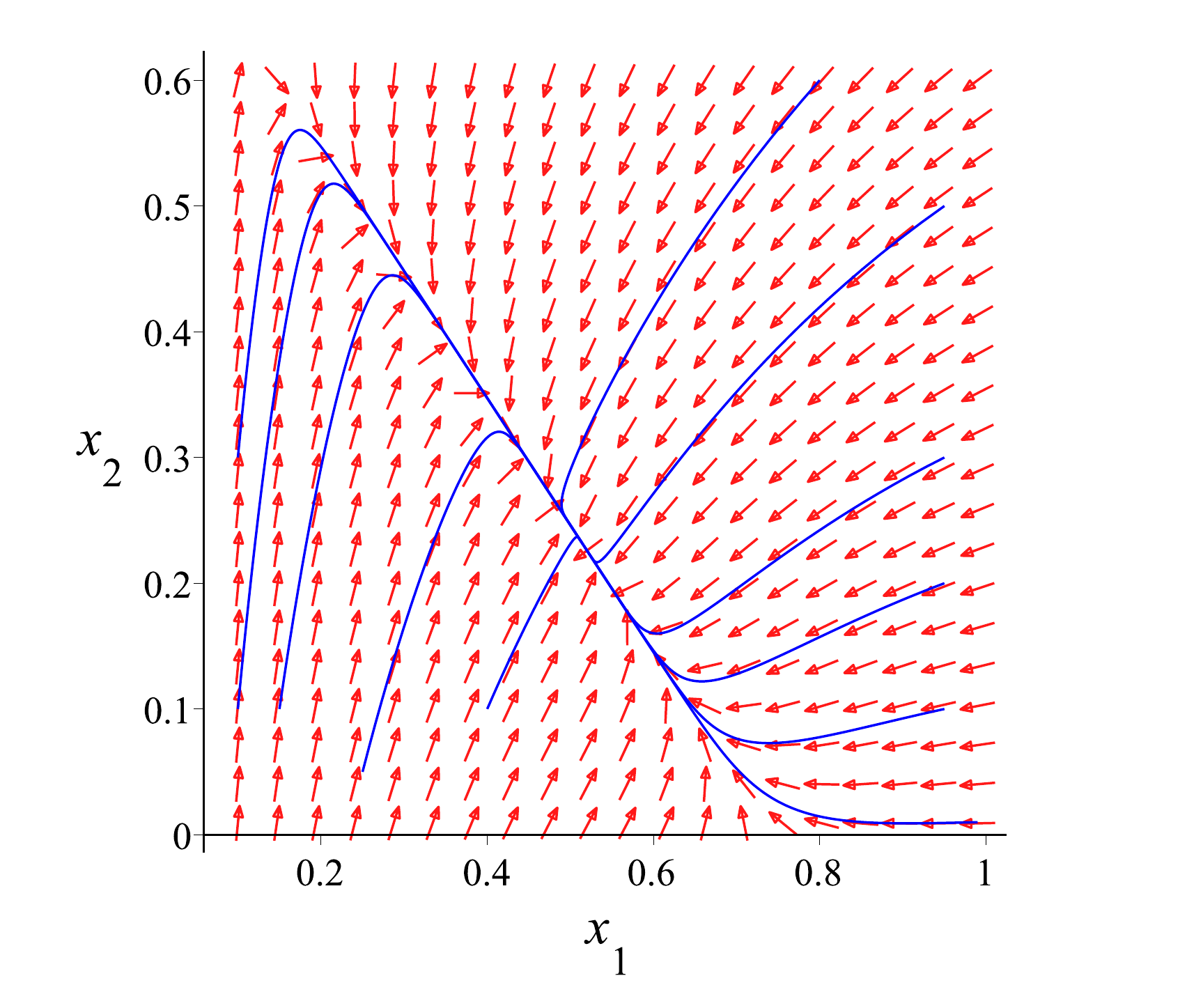, height=170pt,
width=170pt,angle=0}
\end{center}
\caption{The phase portrait of average populations for (a) $r_{1} =1.0, r_{2} = 0.8, u_{1,2}=0.5, v_{1,2}=0.0, d_{1,2,\rm D}=1.0$ with dedifferentiation rate $q = 0.0$. (b) Same parameter values as part (a) but $q = 0.5$ now. The important change here is that the fixed point $x^{*}_{1}$ remains along the horizontal axis, indicating the possibility of extinction of mutant cells, while
the new fixed point (previously called $x^{*}_{2}$ now have non-zero values of normal and mutant stem cells indicating the possibility of co-existence between the two population in the presence of dedifferentiation. Notice the co-existence can only occur when the disadvantageous mutant becomes advantageous in the presence of dedifferentiation).}
\label{portrait34}
\end{figure}

We have depicted the phase portrait of the stem cell dynamics in Fig. \ref{portrait34}. While the vertical fixed point of mutant stem cells is not globally stable for $r_{2}/r_{1} = 0.8$, upon introduction of a finite value of dedifferentiation, the role of the two fixed points switches and the mutant is now advantageous (Fig. \ref{portrait34}b).

In the next section we analyze the numerical solutions of both replicator dynamics in detail. The numerical solutions will let us calculate the time to fixation of the model as a function of proliferation and differentiation rates of normal and cancer stem cells as well as death rates of all three populations.

\section{Numerical Results}
\subsection{Time to Fixation and Steady State Fraction of Cancer Stem Cell}

\begin{figure}
\begin{center}
\epsfig{figure=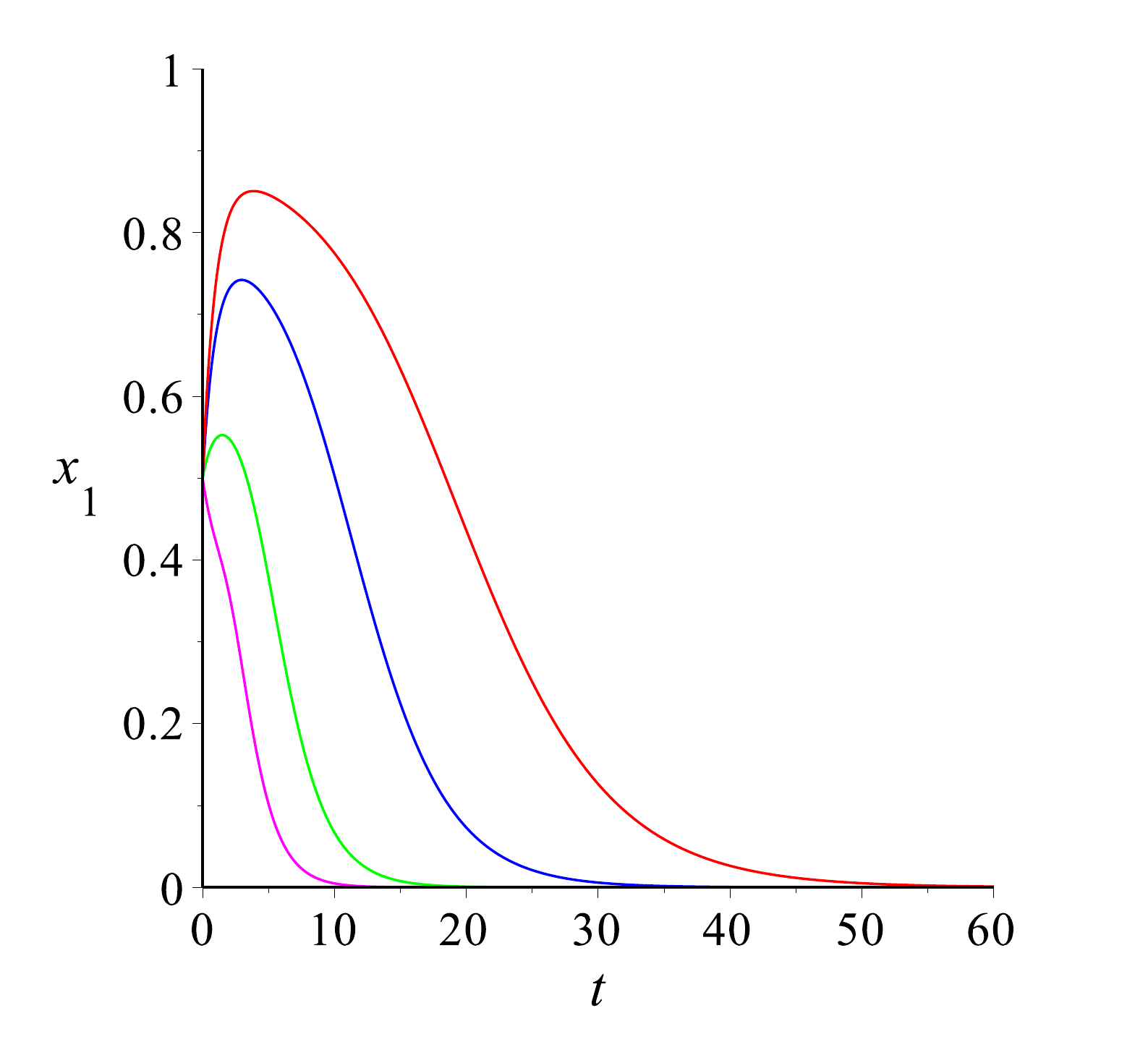, height=150pt,
width=150pt,angle=0}
\epsfig{figure=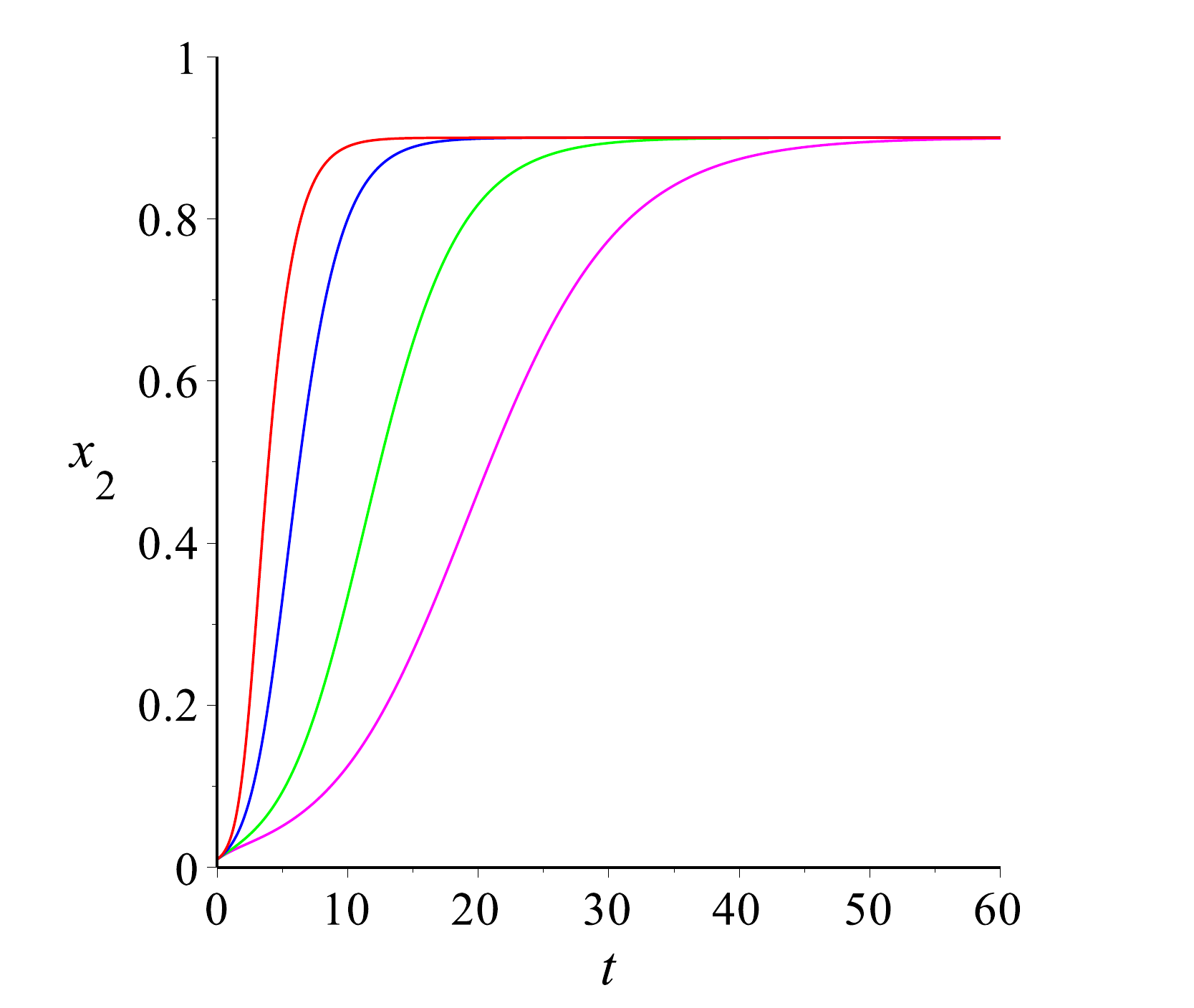, height=150pt,
width=150pt,angle=0}
\epsfig{figure=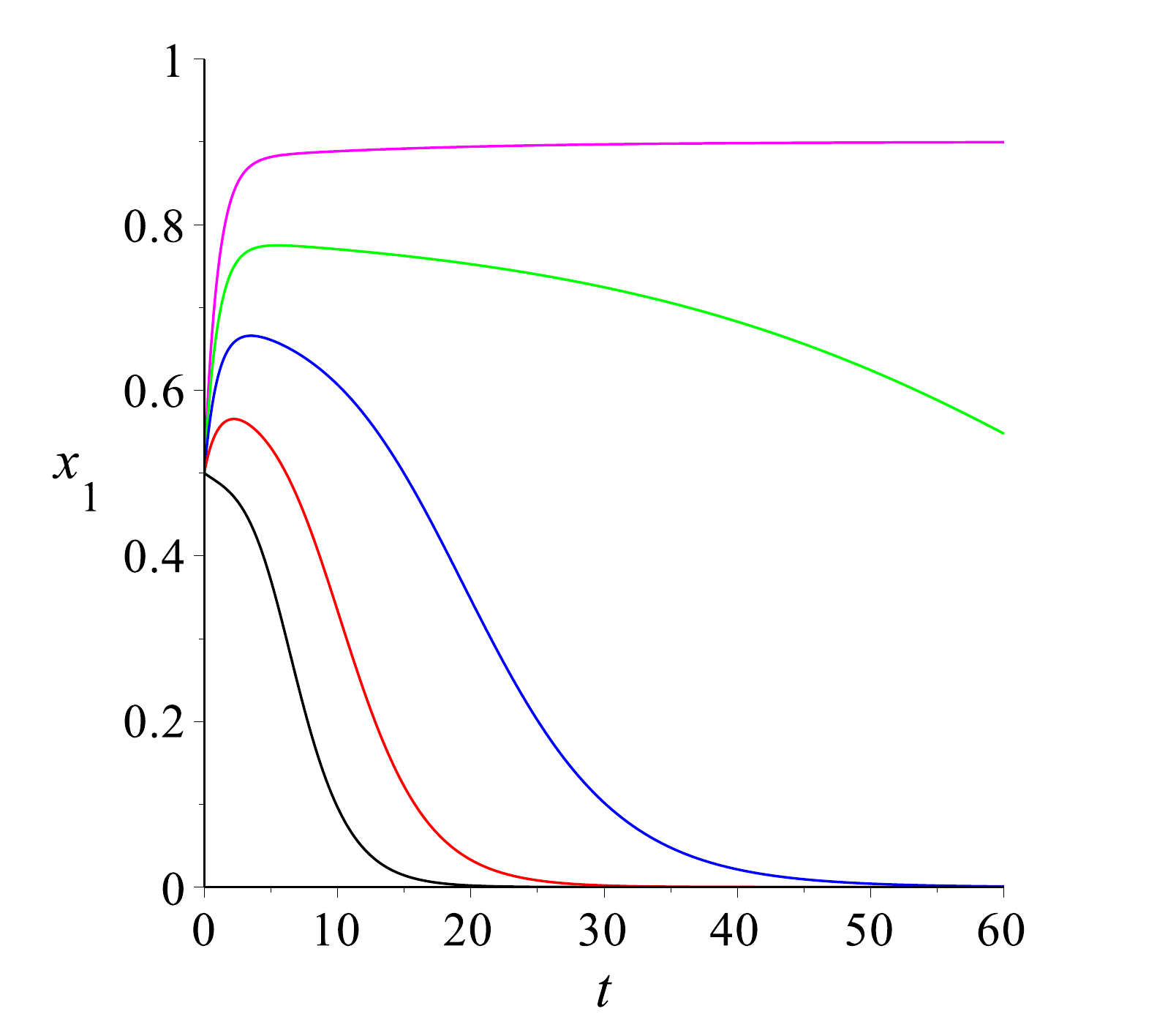, height=150pt,
width=150pt,angle=0}
\epsfig{figure=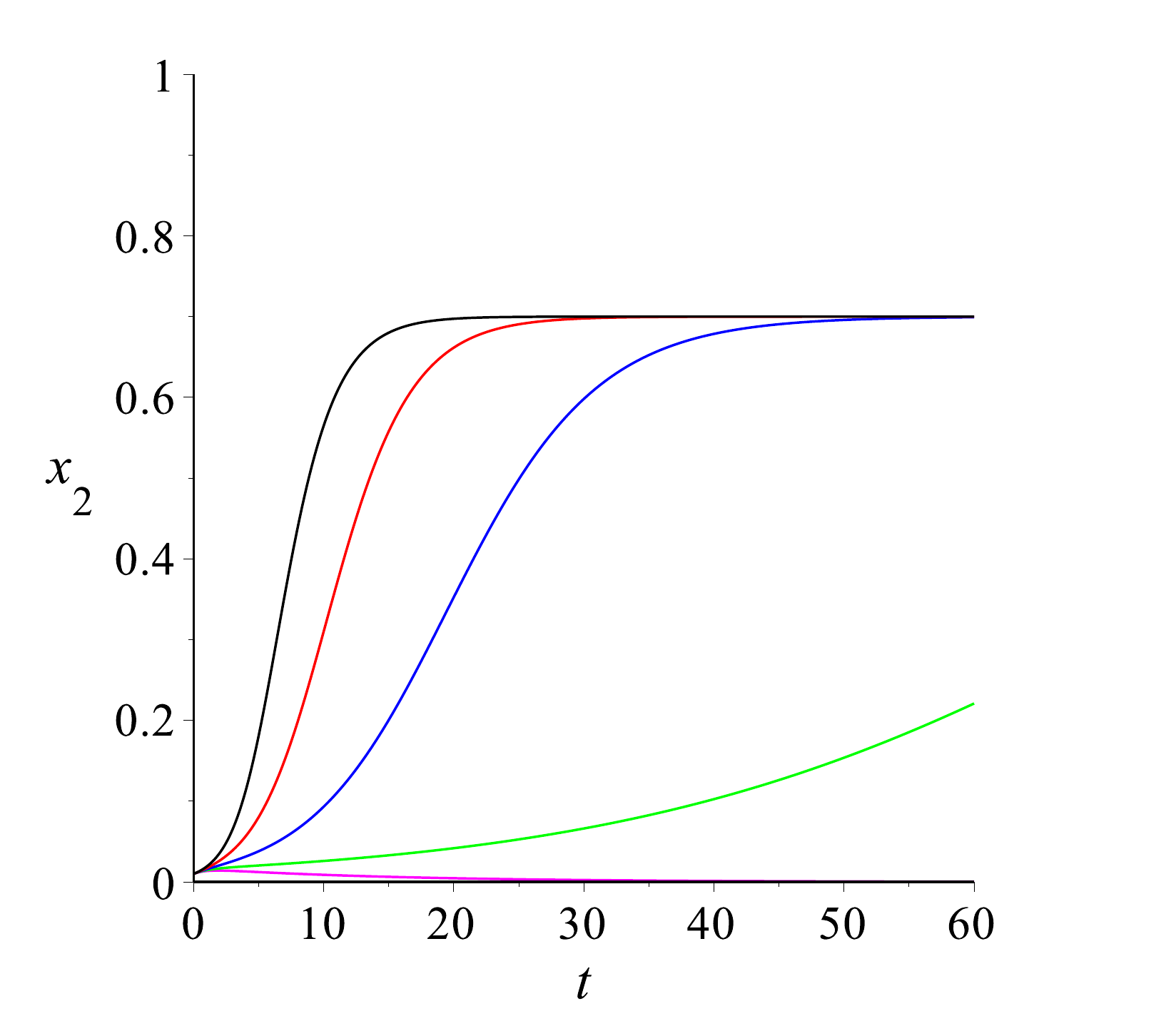, height=150pt,
width=150pt,angle=0}
\end{center}
\caption{ Stem cell frequencies as a function of time for various division rates and differentiation probabilities: (top)$r_{1} = 1, r_{2}=1.2,  u_{1} = 0.2$ (magenta), 0.4 (green), 0.6 (blue), 0.8 (red), $u_{2} = 0.1, v_{1,2}=0.0, d_{1,2,\rm D}=1.0$. ( bottom) $r_{1} = 1, r_{2}=1.2,  u_{1} = 0.1$ (magenta), 0.2 (green), 0.3 (green), 0.3 (blue), 0.4 (red), 0.5 (black), $u_{2} = 0.3, v_{1,2}=0.0, d_{1,2,\rm D} = 1.0$.}
\label{fix-diff-no11-12}
\end{figure}

We numerically integrate the system of equations in Eq. \ref{replicator} and plot (Fig. \ref{fix-diff-no11-12}) the stem cell frequencies as function of proliferation and death rates in addition to differentiation probabilities. In addition to time to fixation, the value of the steady state fractions can be read off from these plots. In most of the cancer initiation cases, the population of normal and differentiated cells are in equilibrium prior to introduction of the mutant stem cell. The equilibrium fraction between stem cells and differentiated cells differ from tissue to tissue but
normally stem cells constitute a small minority. This is reflected in our choice of initial conditions. The initial frequency of normal stem cells is taken to be 0.5 while the mutant frequency is 0.01. In Fig. \ref{fix-diff-no11-12}a proliferation parameters read as $r_{2} = 1.2, r_{1}=1.0, d_{1,2,\rm D} = 1.0$. As it is believed that most of the differentiation events occur asymmetrically, i.e., $v_{1,2} \ll 1$, we choose $v_{1,2} =0.0$ and look at the population dynamics of both stem cell species as $u_{1}$ is varied for two different values of $u_{2}$ = 0.1, Fig. \ref{fix-diff-no11-12} (top), and $u_{2}$ = 0.3, Fig. \ref{fix-diff-no11-12} (bottom). The mutant stem cell is always assumed advantageous with $r_{2}/r_{1} = 1.2$. Notice, while relative fitness in the absence of differentiation is bigger than unity, for low differentiation rates of normal cells and high differentiation rates of cancer stem cells
Fig. \ref{fix-diff-no11-12} ($u_{1}=0.1, u_{2}=0.3$) the normal stem cells have a selective advantage. Also the increase in time to fixation is apparent as one decreases the differentiation rate of normal stem cells. For the normal stem cell frequency, there is a peak at early times
which begins to decrease monotonically. As discussed in the previous section, this early increase in the population of both stem cells indicates that there is an early competition between stem cells and differentiated cells before differentiated cells reach saturation and stem cells begin to compete with each other. As the values of differentiation probabilities is decreased  we observe an increase in time to fixation (time to reach plateau in Fig. \ref{fix-diff-no11-12}) . This is in accordance with the exact analytical result for time to fixation $t_{\rm F}$, Eq. \ref{t-fix}.

\begin{figure}
\begin{center}
\epsfig{figure=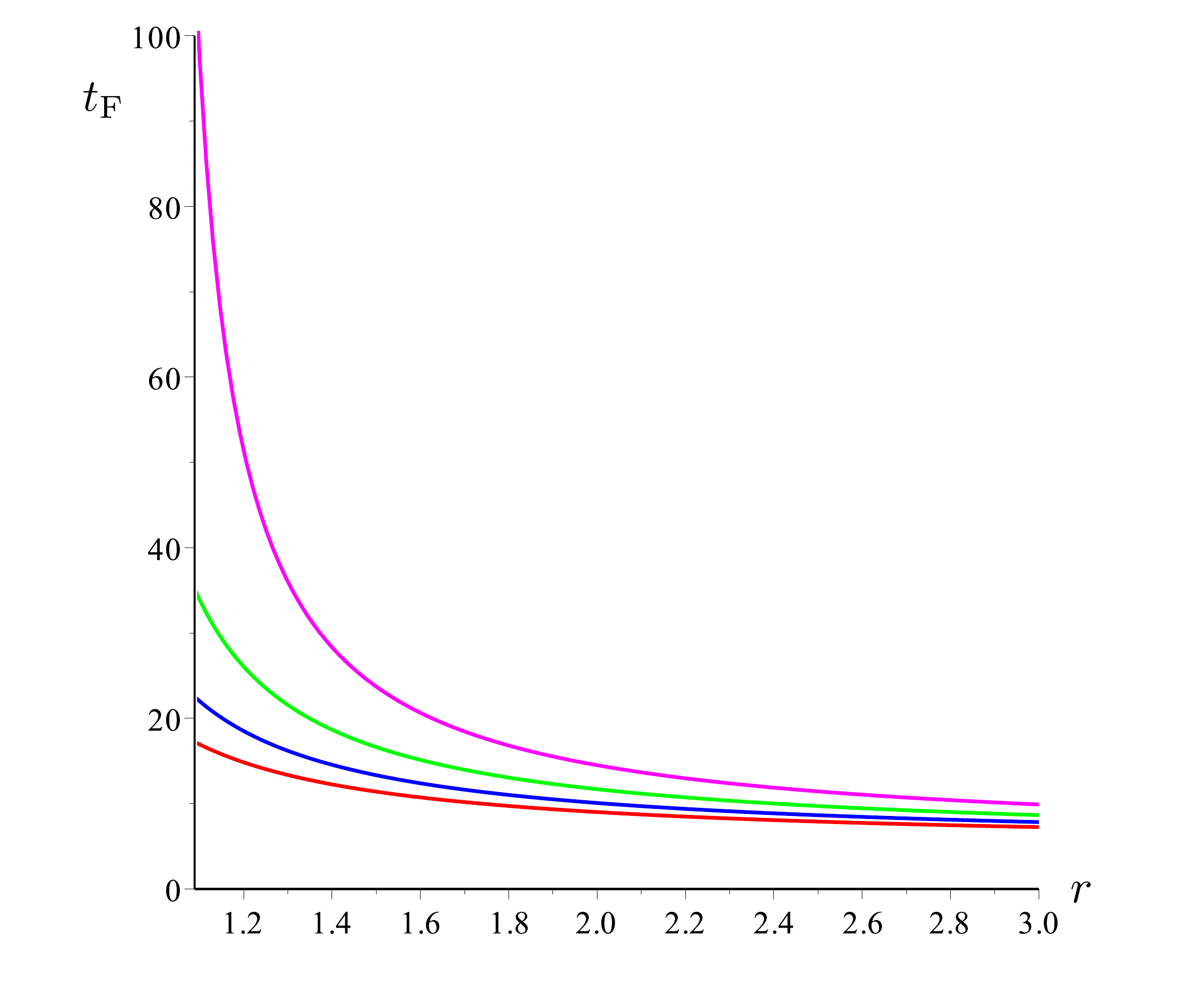, height=150pt,
width=150pt,angle=0}
\epsfig{figure=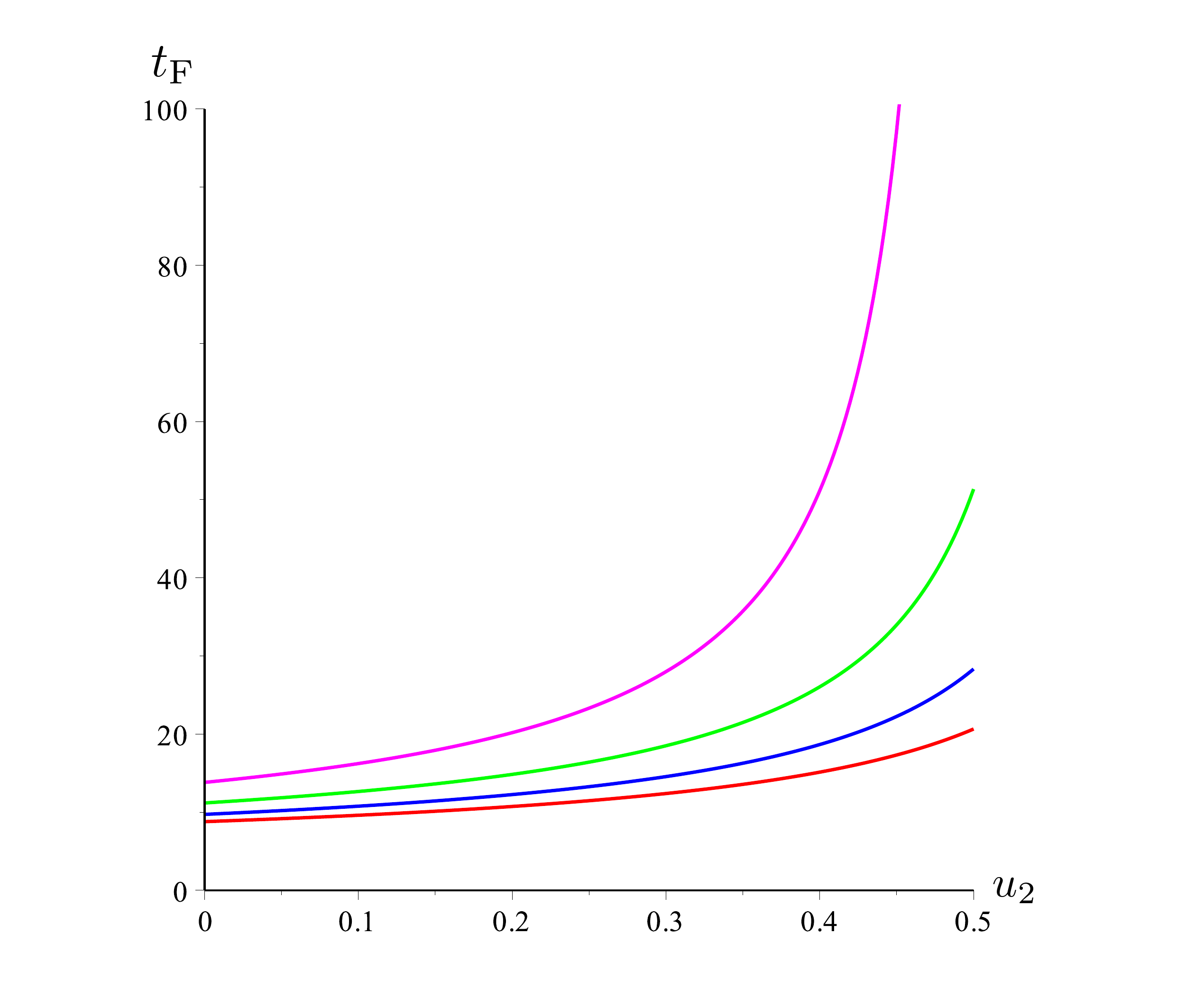, height=150pt,
width=150pt,angle=0}
\end{center}
\caption{(Average) time to fixation as a function of relative fitness and differentiation rates left: $t_{\rm F}$ as a function of $r_{2}/r_{1}$ for $u_{1}=0.5$ and $u_{2}$=0.2 (magenta),0.3 (green), 0.4 (blue), 0.5 (red). Times are measured in units of generation.}
\label{tfix-plot}
\end{figure}

Time to fixation is also plotted directly from integration of Eq. \ref{replicator} as a function of both relative fitness and differentiation rates. This is plotted in Fig. \ref{tfix-plot}. Again, as is expected from Eq. \ref{t-fix}, time to fixation dramatically reduces as division ratio $r_{2}/r_{1}$ increases as a function of asymmetric differentiation rate of cancer stem cells.

\begin{figure}
\begin{center}
\epsfig{figure=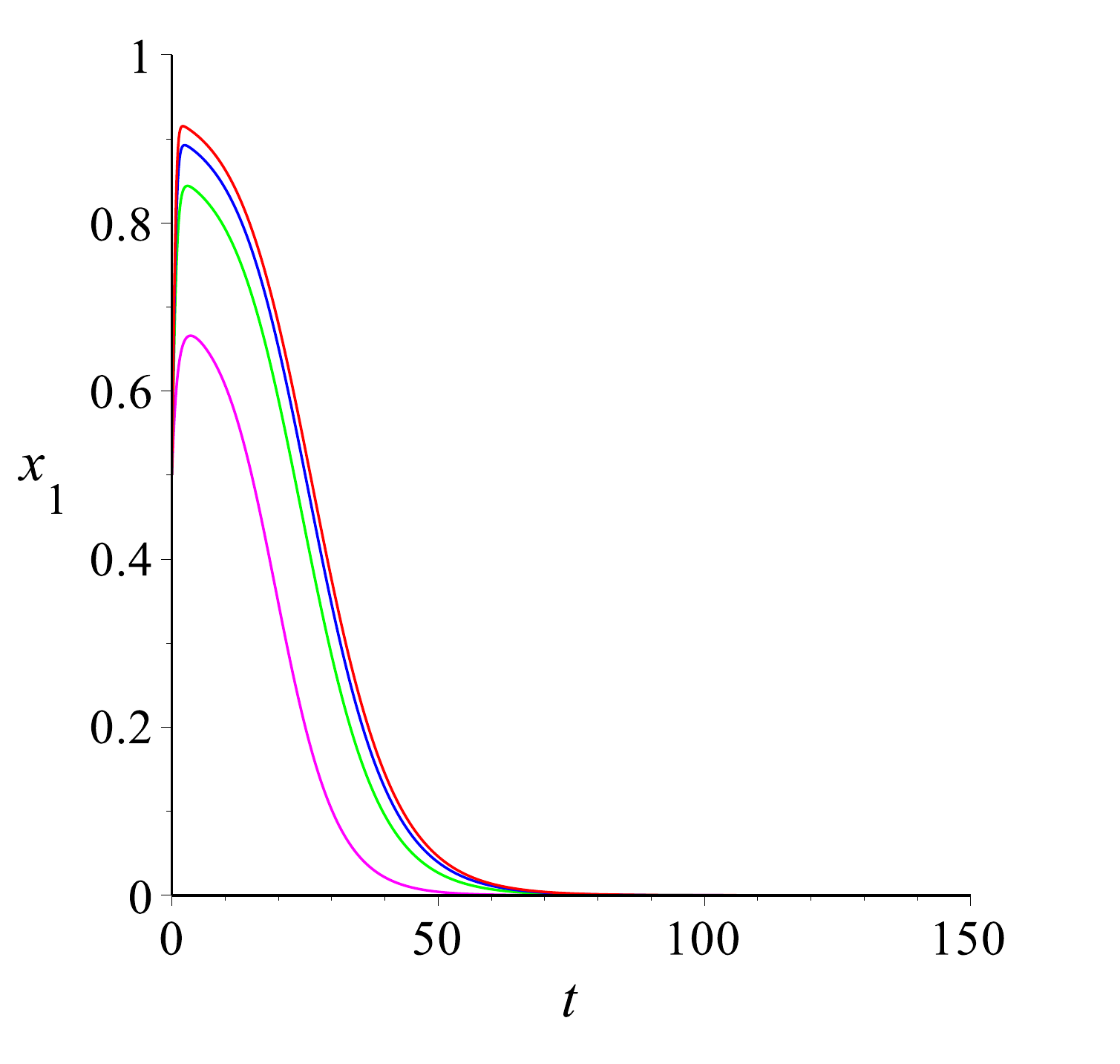, height=150pt,
width=150pt,angle=0}
\epsfig{figure=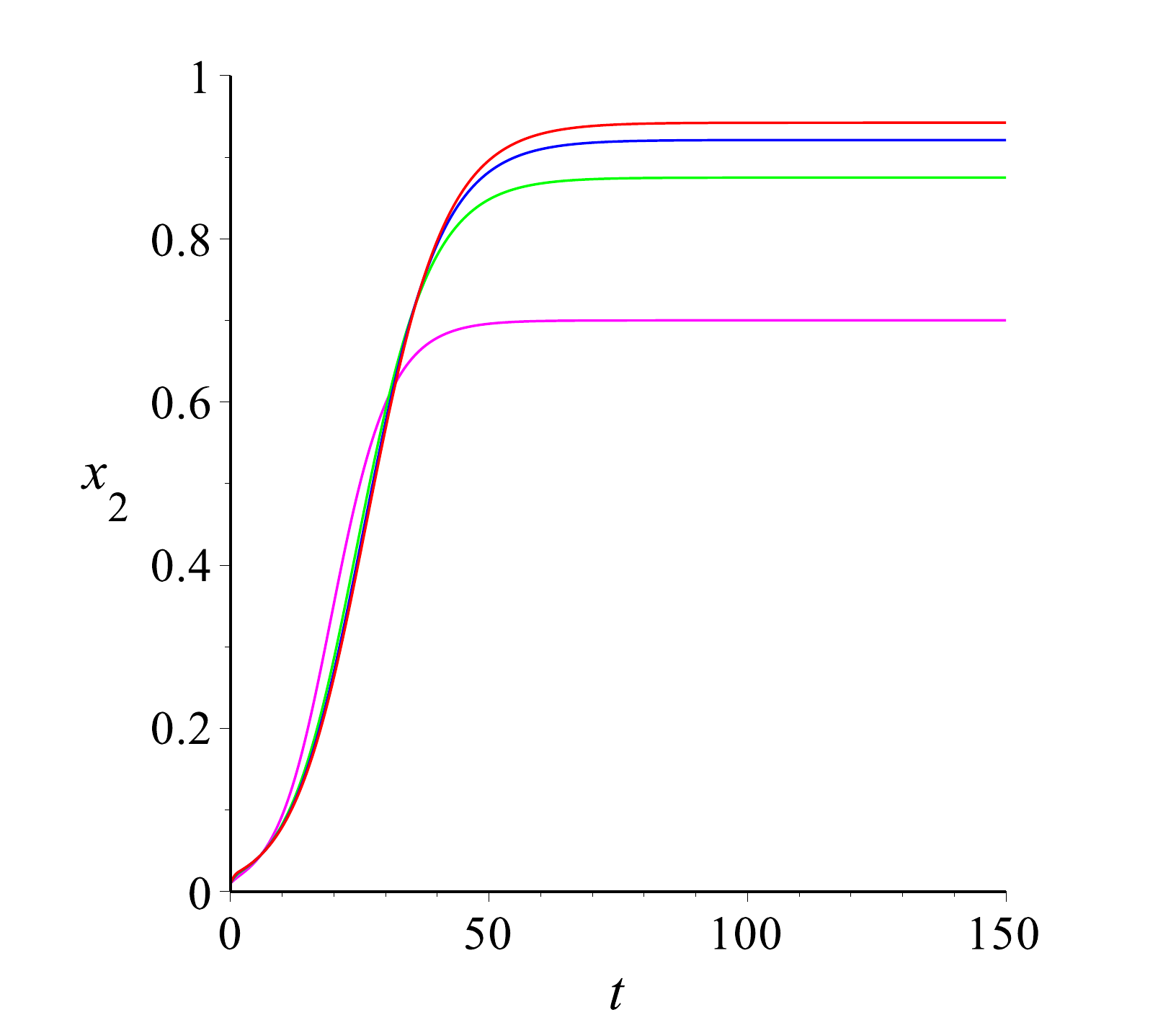, height=150pt,
width=150pt,angle=0}
\end{center}
\caption{Stem cell fractions for parameters $r_{1} = 1, r_{2}=1.2,  u_{1,2}=0.3, v_{1,2}=0.0, d_{1,2}=1.0, d_{\rm D}$=1.0 (magenta), 3.0 (green), 5.0 (blue), 7.0 (red).}
\label{stem-dD}
\end{figure}

\begin{figure}
\begin{center}
\epsfig{figure=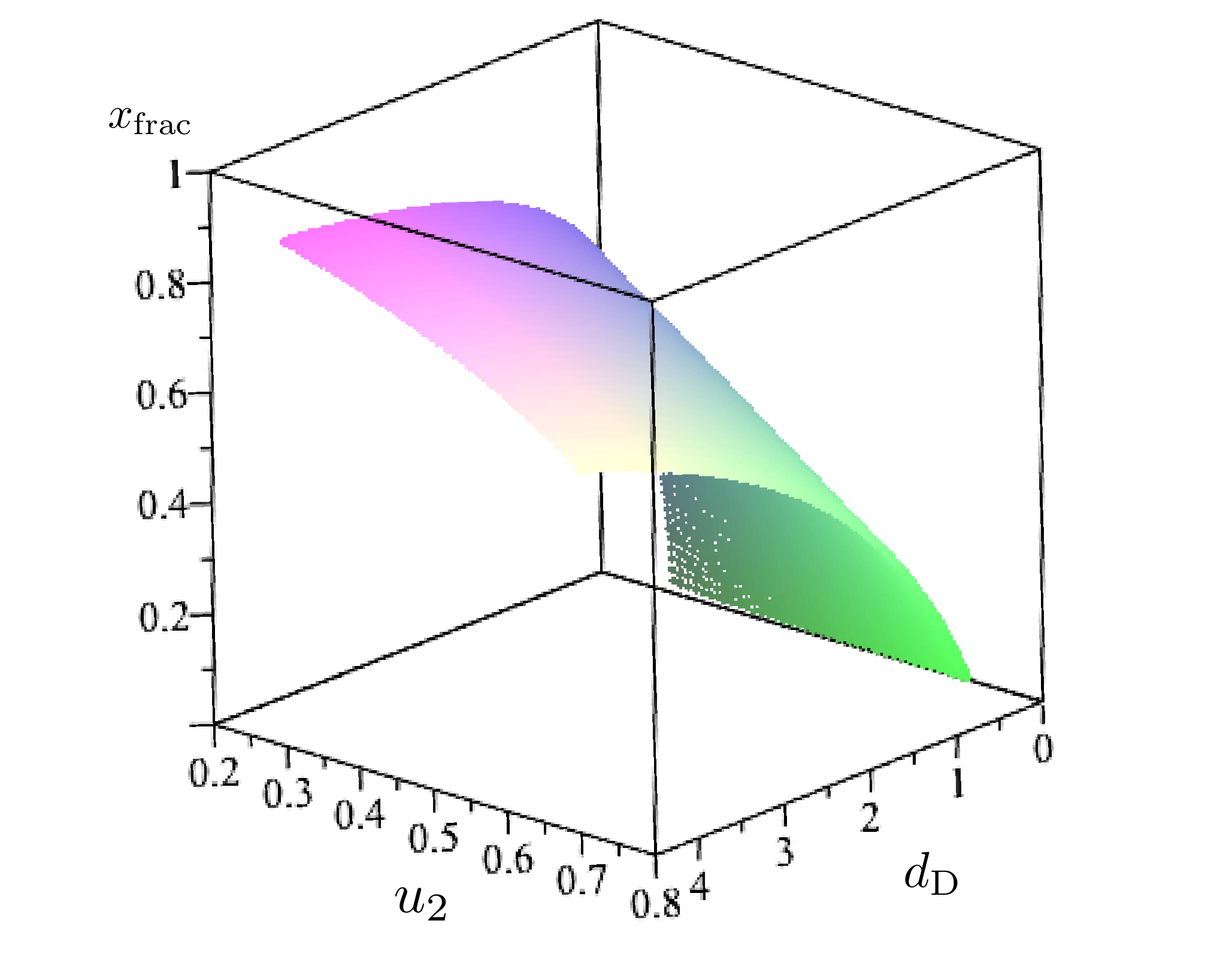, height=200pt,
width=240pt,angle=0}
\end{center}
\caption{Mutant stem cell fraction in the steady state as a function of differentiation rate $u_{2}$ ($u_{1}=u_{2}, v_{1,2}=0.0$) and the
death rate of differentiated cells, $d_{\rm D}$. Notice that the limit of low death-rate of differentiated cells is not a physical result.
In fact, in tissues such as intestinal crypt, the differentiated cell death rates are significantly higher than stem cells.}
\label{stem-u2dD}
\end{figure}

As discussed in the previous section, the effective fitness is in fact independent of the value of $d_{\rm D}$ as can be seen from the condition in Eq. \ref{replicator-fitness}. However, changing $d_{\rm D}$ affects the final frequency of mutant stem cells. This can be seen in Fig. \ref{stem-dD}. Fig. \ref{stem-dD}a is plotted with parameter values $r_{1} = 1, r_{2}=1.2,  u_{1,2}=0.3, v_{1,2}=0.0, d_{1,2}=1.0, d_{\rm D}$=1.0 (magenta), 3.0 (green), 5.0 (blue), 7.0 (red). Finally, we plot the fraction of mutant stem cells as a function of both $d_{\rm D}$ and the asymmetric differentiation rate of mutant cells, $u_{2}$ in Fig. \ref{stem-u2dD}.

\subsection{Cancer Stem Cell Dynamics in the Presence of Dedifferentiation}

\begin{figure}
\begin{center}
\epsfig{figure=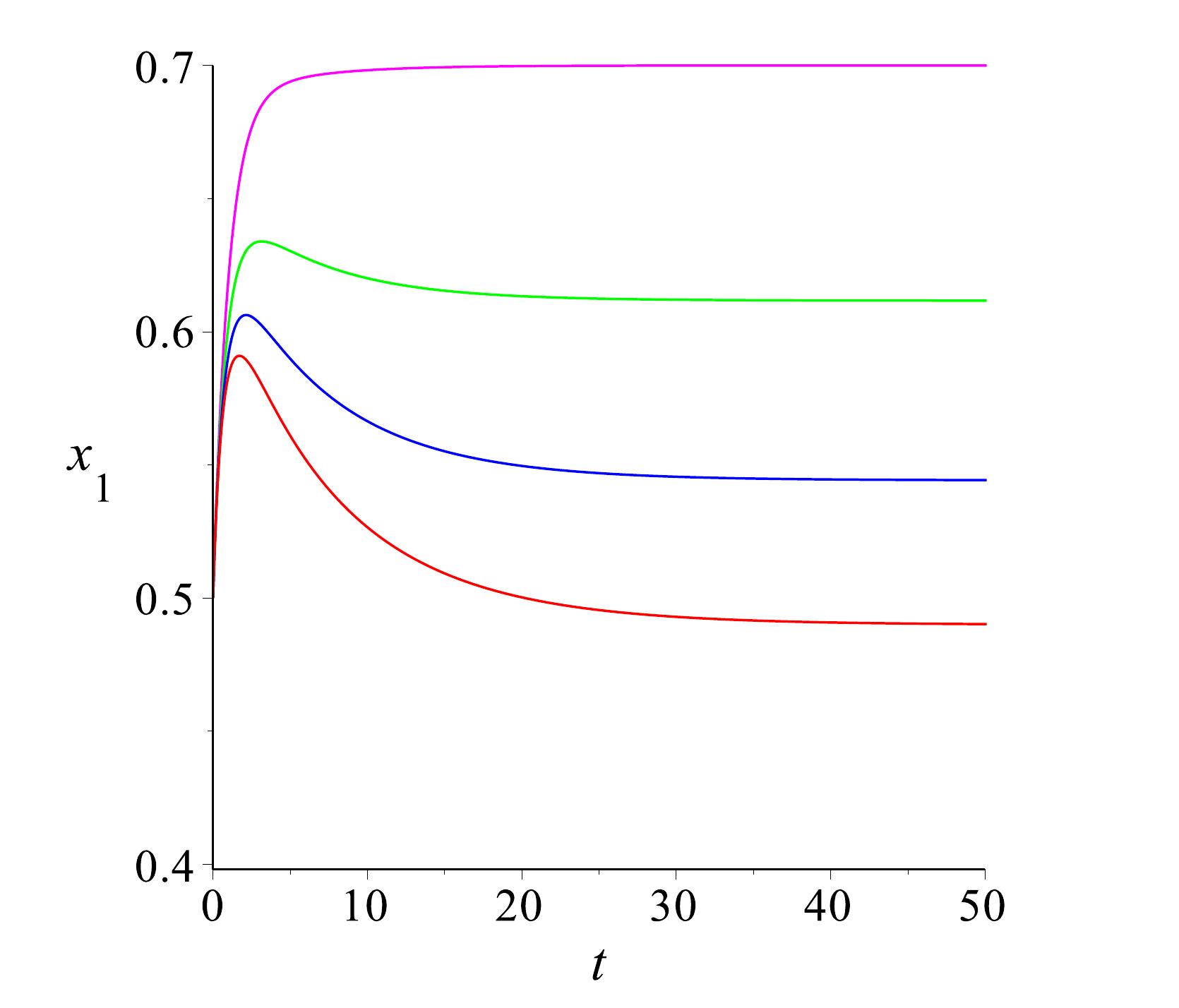, height=150pt,
width=150pt,angle=0}
\epsfig{figure=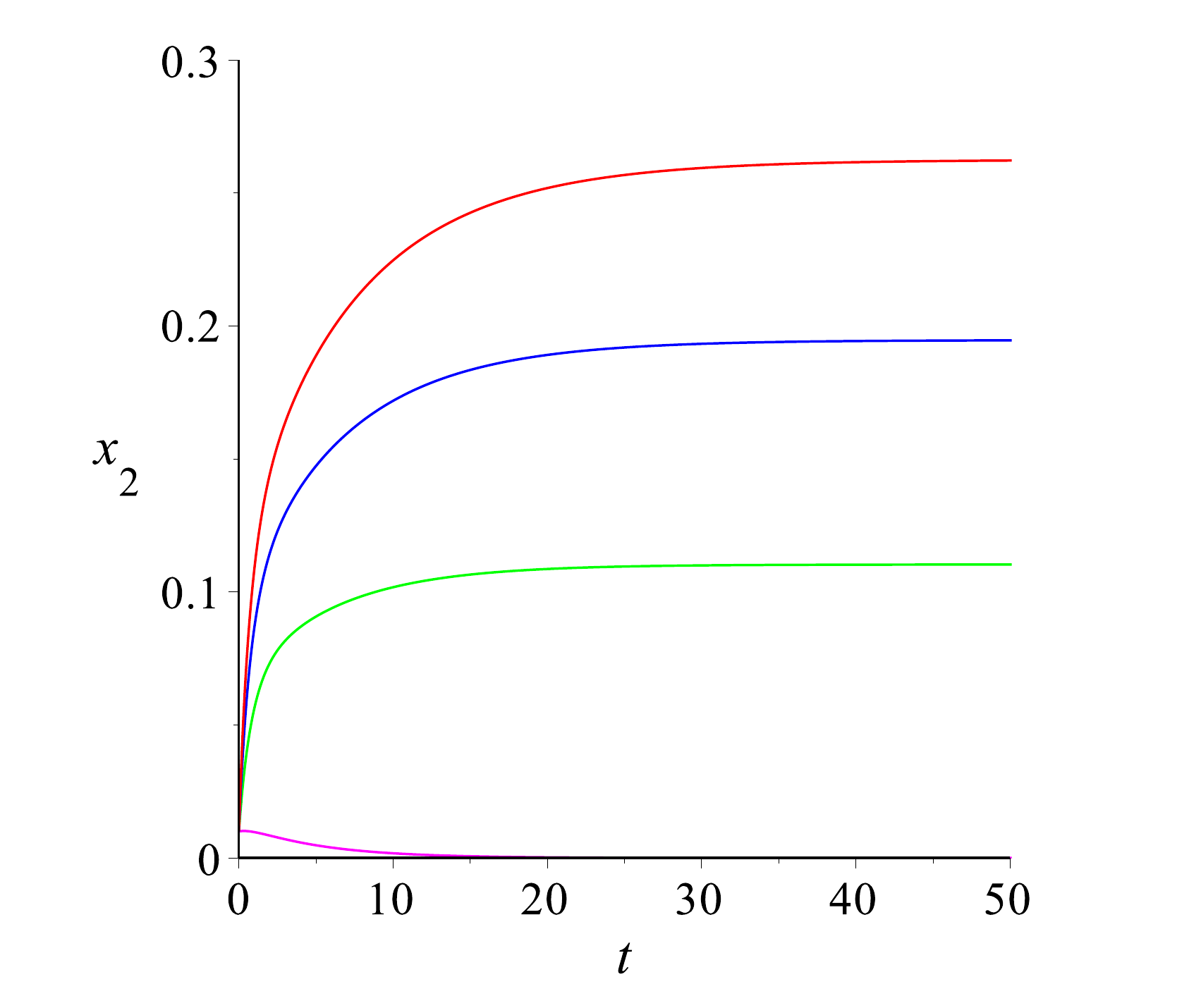, height=150pt,
width=150pt,angle=0}
\epsfig{figure=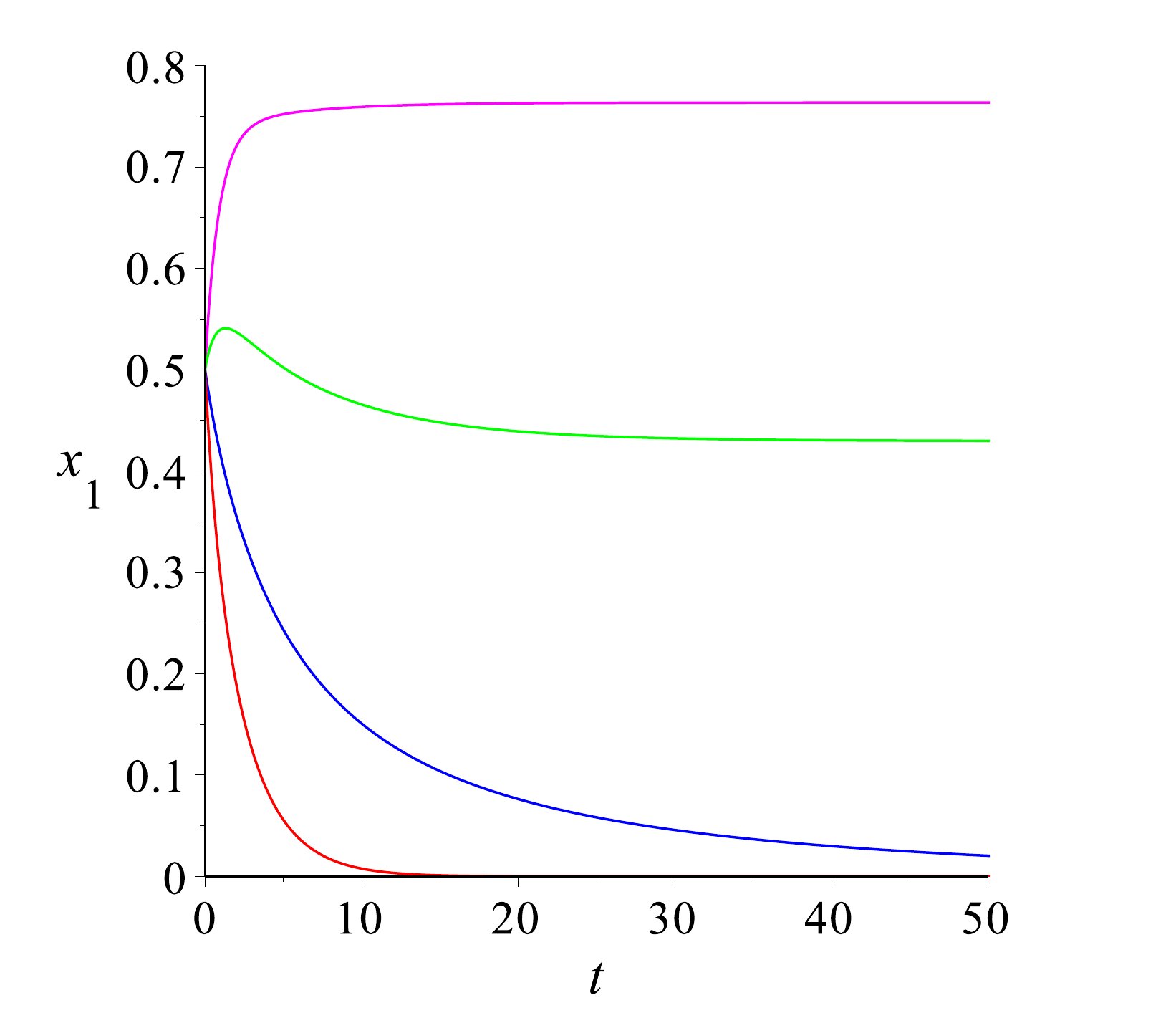, height=150pt,
width=150pt,angle=0}
\epsfig{figure=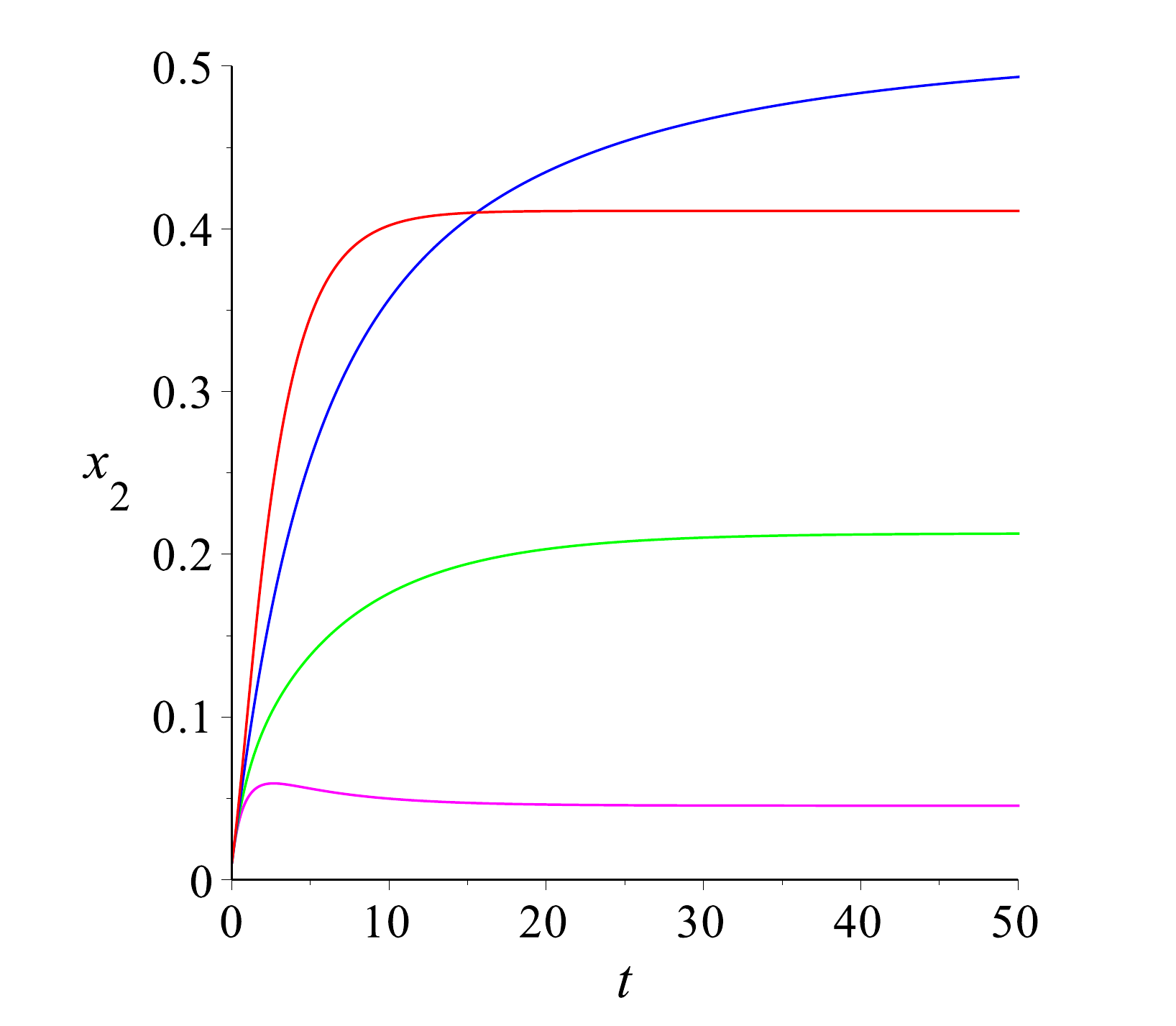, height=150pt,
width=150pt,angle=0}
\end{center}
\caption{Stem cell dynamics in the presence of differentiation. Top: Dynamics of both mutant and normal stem cell frequencies for different values of dedifferentiation rate $q$, while rest of parameters are kept constant at $r_{1} = 1, r_{2}=0.8,  u_{1,2} = 0.3, v_{1,2}= 0.0, q=0.0$ (magenta), $q=0.2$ (green), $q=0.4$ (blue), $q=0.6$ (red). Bottom: Dynamics of populations as differentiation rate varies: $r_{1} = 1, r_{2}=0.8, v_{1,2}=0.0, q= 0.2$, $u_{1,2}=0.2$ (magenta), $u_{1,2}=0.4$ (green), $u_{1,2}=0.6$ (blue), $u_{1,2}=0.8$ (red). }
\label{stem-dediff}
\end{figure}

In the presence of dedifferentiation Eq. \ref{plasticity} can be solved for different values of dedifferentiation and proliferation rates of normal and cancer stem cells. Population dynamics of stem cell frequencies is plotted (Fig. \ref{stem-dediff}) as differentiation rates change.  In Fig. \ref{stem-dediff} (top) a disadvantageous mutant with $r_{2}/r_{1}$ = 0.8 is assumed. As one can see, upon increasing the dedifferentiation rate from 0 to 0.6, the mutant invasive fate changes. For $q=0.2, 0.4, 0.6$, as a matter of fact, the mutant is advantageous. Change of differentiation rates also play an important role in the dedifferentiation dynamics. Dedifferentiation not only depends on the rate of conversion and phenotype switching of non-stem cells but also on the population of non-stem cells itself.
Based on our model this is not surprising that as the differentiation rate $u_{1,2}$ (in this case) increases the originally disadvantageous cancer stem cell becomes advantageous, Fig. \ref{stem-dediff}(bottom).

\begin{figure}
\begin{center}
\epsfig{figure=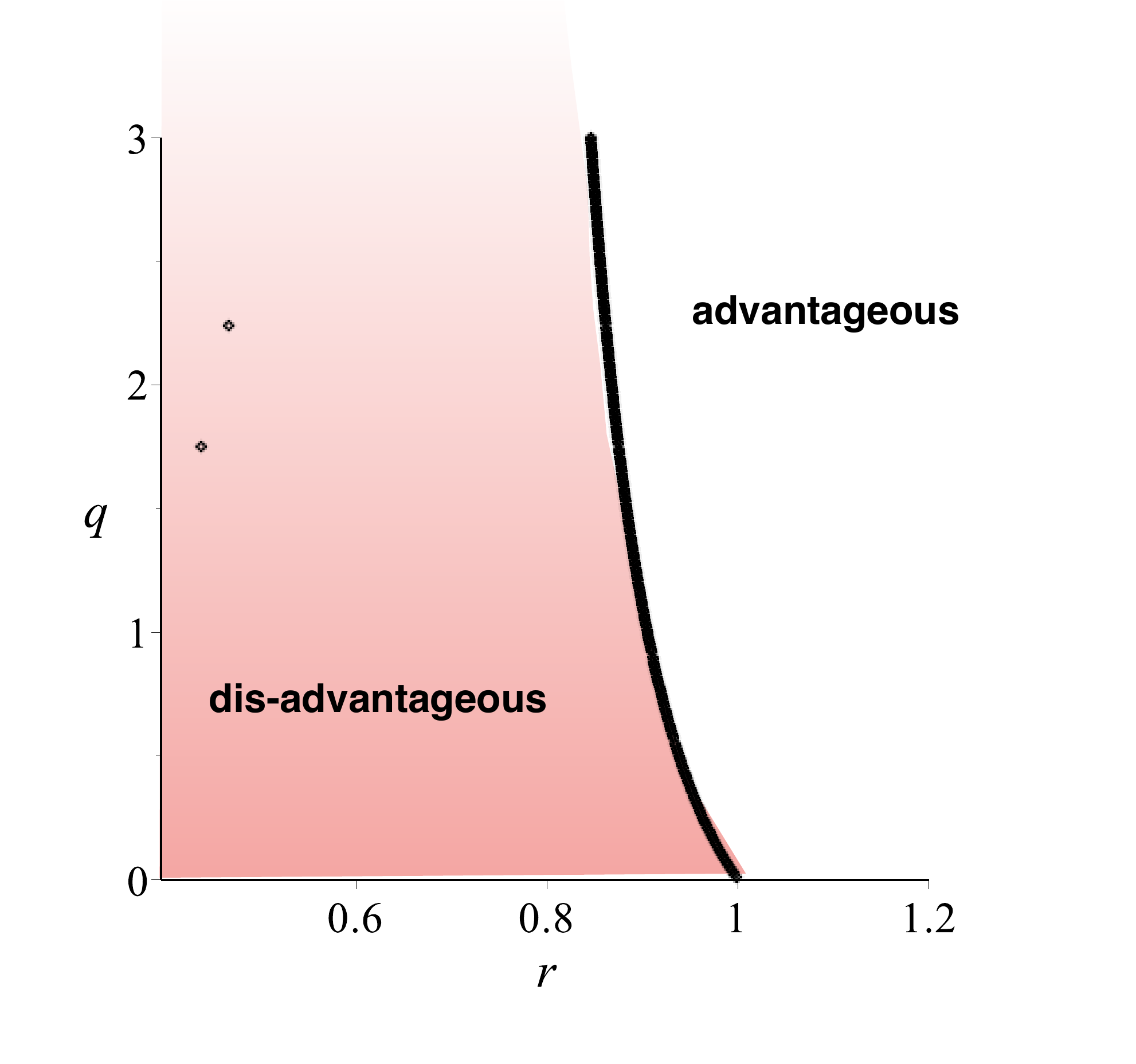, height=170pt,
width=170pt,angle=0}
\epsfig{figure=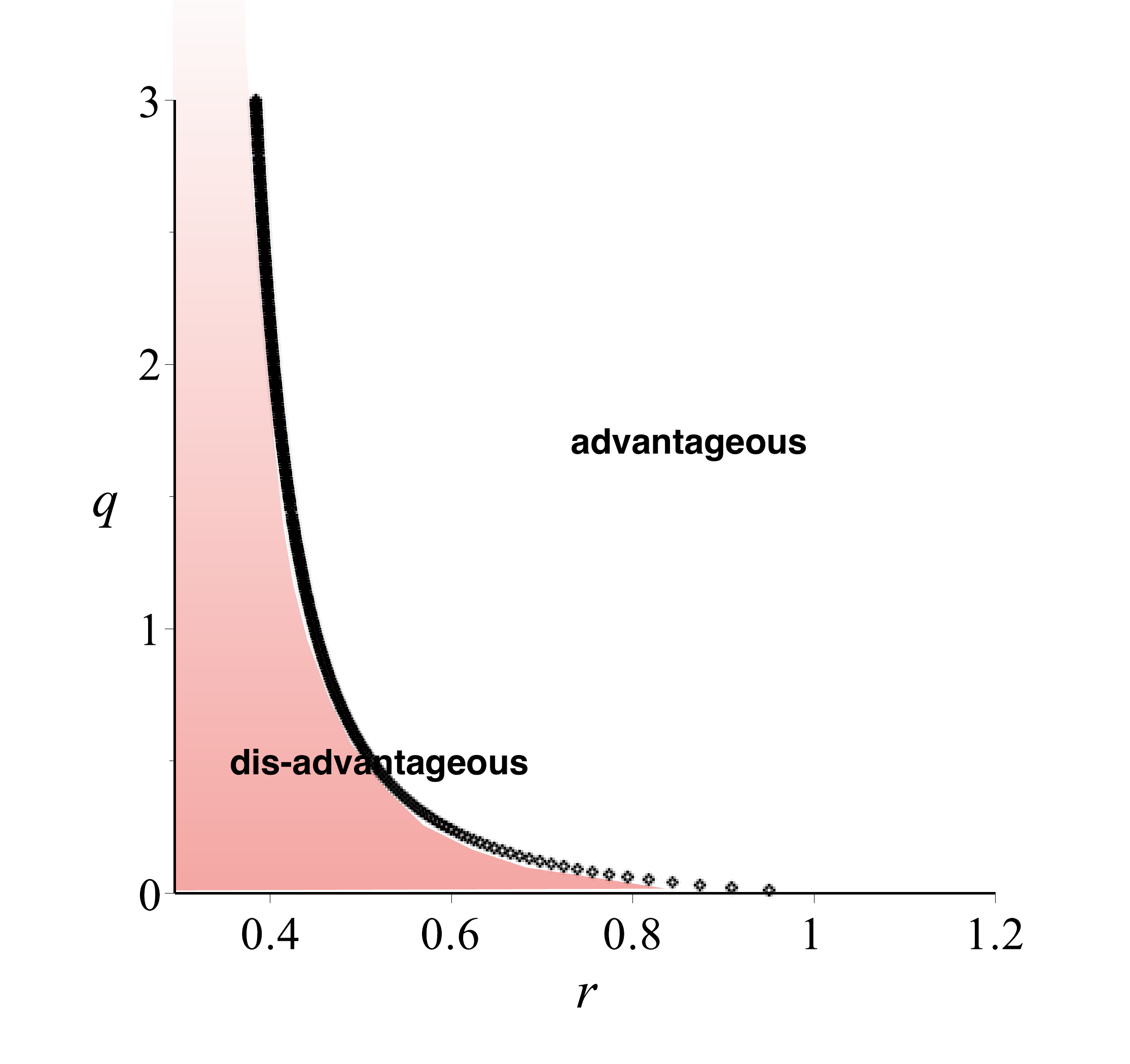, height=170pt,
width=170pt,angle=0}
\end{center}
\caption{Phase diagram of dis-advantageous/advantageous mutant in the space of proliferation rate ($r_{2}$ represents relative proliferation rate as $r_{1}=1.0$). Three different plots correspond to different value of differentiation probabilities.}
\label{phase-dediff12}
\end{figure}

\begin{figure}
\begin{center}
\epsfig{figure=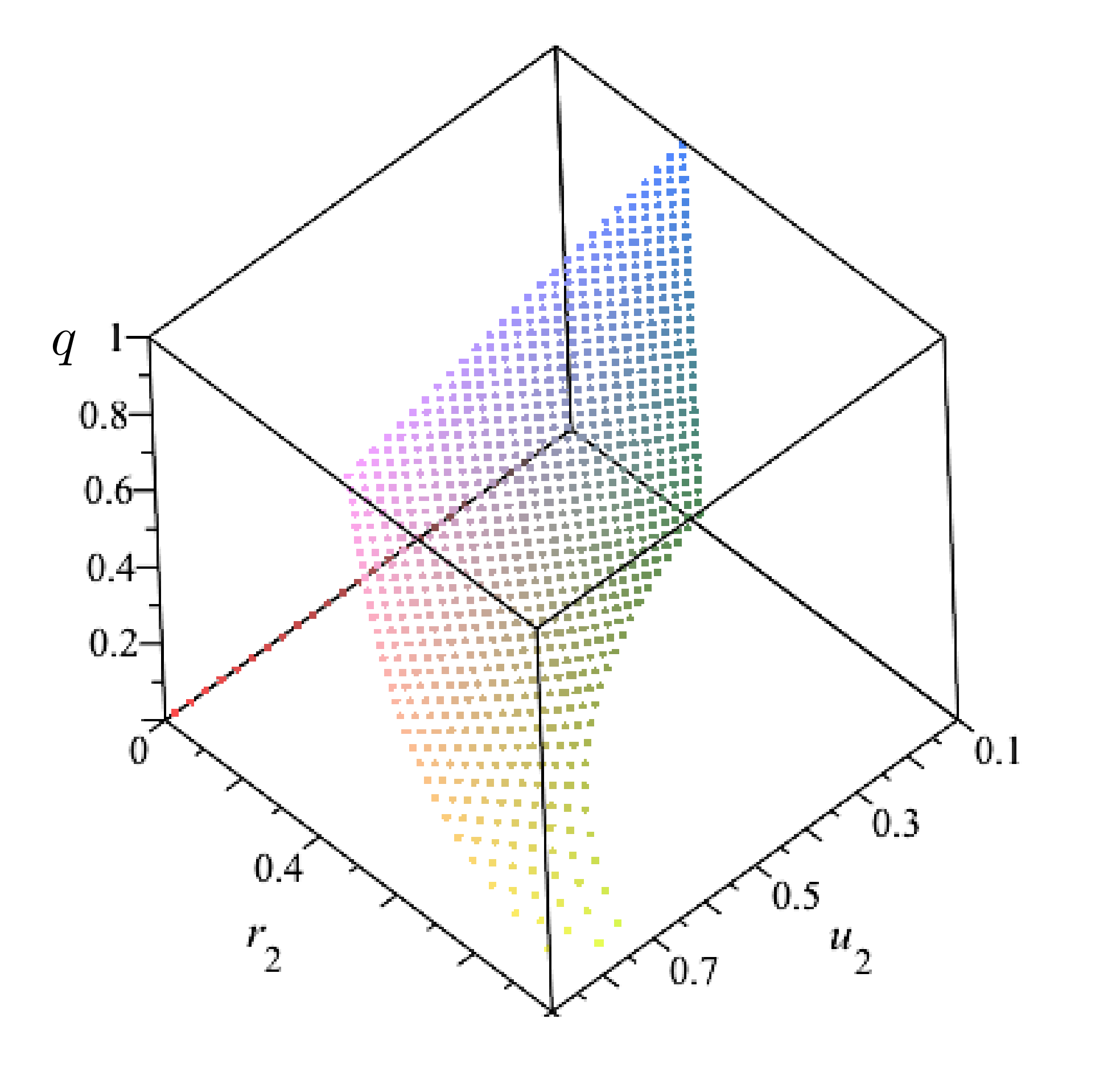, height=250pt,
width=250pt,angle=0}
\end{center}
\caption{Three dimensional phase-diagram for dedifferentiation rate of cancer stem cells $q$, proliferation rate $r_{2}$ ($r_{1} = 1.0$) and asymmetric differentiation probability $u_{2}$. ($u_{2} = u_{1}, v_{1,2}=0.0$). The curve indicate the neutral limit.  All points in the phase space in the left belong to an dis-advantageous mutant case. }
\label{phase-dediff-3D}
\end{figure}

Condition under which the cancer stem cells (in presence of a finite differentiation rate ($u_{1,2}$), dedifferentiation rate ($q$) and relative proliferation ($r_{2}/r_{1}$)) are advantageous or not, is a nontrivial question. As can be seen from the fixed-point analysis, the two fixed points corresponding to a full fixation of mutant cell (versus partial) in the final population always exist. The condition for having a successful mutant depends on when the mutant fixed point becomes globally attractive. We have calculated this numerically by diagonalizing the Jacobian matrix of Eq. \ref{plasticity}, and looking at the point when the second eigenvalue becomes negative (one eigenvalue corresponding to $x_{2}$ direction is always attractive thus we only need to look at the sign change of the other eigenvalue).This leads to a phase diagram for successful and unsuccessful mutants as function of $r= r_{2}/r_{1}$ and $q$. Fig. \ref{phase-dediff12} shows the results for two different values of differentiation rates $u_{1,2}$= 0.3 (left) and $u_{1,2}$ = 0.7 (right). The higher differentiation rate helps the mutant to shift the neutral critical point (i.e. $r=1$ for $q=0$) faster to lower $r$ as $q$ increases. Fig. \ref{phase-dediff-3D} shows a 3D plot of such a phase diagram as all the three relevant parameters in the model change.

\section{Discussion and Implications on Cancer Therapeutics}

In this work we constructed a general model of stem cell selection dynamics by including the population of differentiated cells into a Moran-type model which governs the stochastic dynamics of both stem cell and non-stem cell species at the same time. Although the non-stem cell (differentiated) population lacks proliferative potential its population can increase through the differentiation of stem cells. Our model allows us to proceed with investigating the conditions for having a successful mutant stem cell as a function of differentiation probabilities as well as the proliferation and death rates. Beside the competition between two stem cells the model now can describe competition between stem cells and non-stem cells which resemble itself as the fraction of the stem cells in the steady state. The replicator dynamics that is derived for stem cell frequencies can predict the average time to fixation which is in agreement with previous known results for the Moran model in the limit of no differentiation.

The above construction allows us to approach theoretically the question of dedifferentiation and phenotypic plasticity that has been observed in recent experiments and which is believed to be an important factor at the latest stages of cancer. By including dedifferentiation events into our replicator dynamics, we are able to determine the fate of a cancer stem cell not only by its proliferation potential but also based on the dedifferentiation rate and overall differentiation parameters. We derive a 3D phase diagram that determines in what regions of (proliferation, dedifferentiation, differentiation) space the cancer (mutant) stem cell is successful. Interestingly, the phenotypic heterogeneity that is believed to be a result of cell plasticity is seen in our model in the presence of dedifferentiation. Due to the plastic nature of differentiated cells even a disadvantageous mutant will not completely go extinct but rather keepsis preserved as a small
subpopulation in the final population of the system and can co-exist with the normal stem cell population.

Notice that due to the complicated nature of adapting hierarchal proliferation scheme of stem cells into constant-finite population model(s) of selection dynamics, we have had to keep things simplified. Thus the population of the differentiated cells generated by either the cancer stem cells or normal stem cells are both included in one compartment. The justification for this assumption, is that in equilibrium there is a relatively large pool of differentiated cells that then due to their plasticity (at least a fraction of them) can be expected to contribute to an influx of
differentiated cells that transform back into caner stem cells. The lack of any experimental observations to distinguish between plastic and non-plastic differentiated cells has led us to keep the simplest model possible. However, the reader should be cautious since for small populations of differentiated cells we might expect some distinction between a lineage of differentiated cells derived from cancer stem cells from one generated by normal stem cells. This will be the subject of future work.

The implications of the cancer stem cell dynamics for cancer therapeutics is also important. The evolutionary dynamics of drug resistance has been discussed in the literature in some detail \citep{key:Iwasa-drug} (see also \citep{key:dingli}\citep{key:foo}). The idea in previous modelling schemes is that due to strong genetic heterogeneity of stem cells
inside the tumour, there is a finite probability of having a particular drug-resistant phenotype which upon chemotherapy (or radiotherapy) can take over the system and replenish the cancer stem cell pool even faster than before. Inclusion of phenotypic plasticity, as we discussed in this model, however, suggests an alternative picture. In this picture tumour heterogeneity is due to diverse dedifferentiation programs which result in a set of plastic differentiated cells transforming back into a variety of cancer stem cell phenotypes. Thus even if the therapy kills all cancer stem cells in the tumour, as long as a small population of plastic non-stem cells exist there is a finite chance for them to create a drug-resistant phenotype. Due to a much higher rate of dedifferentiation (relative to somatic mutations) this alternative picture can explain the solidity of the drug-resistance mechanism in many tumours. Our model also
suggests that this rate strongly depends on the number of dedifferentiation channels developed in the tumour (which correlates with the level of plasticity and aggressiveness of the tumour phenotypes). Drug resistance due to dedifferentiation also predicts a dependence of resistance on the size of the tumour after the therapy. As a matter of fact when the tumour is in its early stages, the dedifferentiation rate is low, the heterogeneity is also small, and thus therapy  eradicates or reduces the size of the tumour significantly. However, if the drug-resistant colony is already formed and the main mechanism of their population increase is due to their self-renewal capability, then the cancer therapy might not work due to both increased rate of plasticity and also inability to overcome the resistant colony.

There are also important implications related to the epithelial-mesenchymal transition (EMT), a cellular differentiation program wherein epithelial cells adopt mesenchymal features. It is known that cancer cells that undergone EMT have less proliferation potential (division rate) \citep{key:gatenby-microenvironment} but much higher plasticity and dedifferentiation than their progenitors. The phase diagram derived in the previous section can answer the paradoxical situation that even though EMT cells are assumed to be disadvantageous inside a tumour they can undergo clonal expansion and defy extinction and eventually pay a significant role in the highly heterogenous population of the tumour.

\section{Appendix A}
\nd Individual events of replication of a normal or mutant stem cell and/or death of a mutant or normal stem cell or a differentiated cell are given by,

\bea
P_{\rm replication,i} &=& \frac{\omega_{i}r_{i}\cdot n_{i}}{r_{1}n_{1} + r_{2}n_{2}}~~~(\omega_{i} \rightarrow u_{i}, v_{i})\nonumber\\
P_{\rm death,k} &=& \frac{d_{k}n_{k}}{d_{1}n_{1}+d_{2}n_{2}+d_{\rm D}n_{\rm D}},~~~i = 1,2~~k = 1,2,D.
\eea

All possible combinations of these replication and death events leads to transition probabilities for the Moran-type model introduced in the text (Fig. \ref{asymmetric}). The master equation for such a three-population death-birth process can be written as,

\bea
N^{-1}\frac{\partial\phi(n_{1}, n_{2}; t)}{\partial t} &=& P^{-,0}(n_{1}+1,n_{2})\phi(n_{1}+1,n_{2};t) + P^{+,0}(n_{1}-1,n_{2})\phi(n_{1}-1,n_{2};t)\nonumber\\
&+&P^{0,-}(n_{1},n_{2}+1)\phi(n_{1},n_{2}+1;t) +P^{0,+}(n_{1},n_{2}-1)\phi(n_{1},n_{2}-1;t)\nonumber\\
&+&P^{-,+}(n_{1}+1,n_{2}-1)\phi(n_{1}+1,n_{2}-1;t) +P^{+,-}(n_{1}-1,n_{2}+1)\phi(n_{1}-1,n_{2}+1;t)\nonumber\\
&+&P^{-,-}(n_{1}+1,n_{2}+1)\phi(n_{1}+1,n_{2}+1;t) + P^{-2,0}(n_{1}+2,n_{2})\phi(n_{1}+2,n_{2};t)\nonumber\\
&+& P^{0,-2}(n_{1},n_{2}+2)\phi(n_{1},n_{2}+2;t) - \{P^{-,0}(n_{1},n_{2}) + P^{+,0}(n_{1},n_{2})\nonumber\\
&+& P^{0,-}(n_{1},n_{2})+P^{0,+}(n_{1},n_{2})+P^{-2,0}(n_{1},n_{2})+P^{0,-2}(n_{1},n_{2}\}\phi(n_{1},n_{2};t),
\eea

\nd where $\phi(n_{1},n_{2};t)$ is the probability density of having $n_{1}$ and $n_{2}$ stem cells (normal and mutant) at time $t$. $P^{\pm, \pm}(n_{1}\pm1, n_{2}\pm 1)$ are transition probabilities to lose or gain one normal or mutant stem cells. $N^{-1}$ is inserted to keep the continuum time limit consistent with discreet time ($\Delta t = 1/N$). Introducing the notation,

\bea
N_{\rm r} &=& r_{1}n_{1} + r_{2}n_{2}\nonumber\\
N_{\rm d} &=& d_{1}n_{1} + n_{2}d_{2} + d_{\rm D}(N-n_{1}-n_{2}),
\eea

\nd the transition probabilities in the absence of dedifferentiation are,

\bea
P^{+,0}(n_{1},n_{2}) &=&  r_{1}\omega_{1}n_{1}\cdot d_{\rm D}(N-n_{1}-n_{2})/(N_{r}N_{\rm d}),\nonumber\\
P^{-,0}(n_{1},n_{2}) &=& \left\{ r_{1}u_{1}n_{1} + r_{2}u_{2}n_{2}\right\}\cdot d_{1}n_{1}/(N_{r}N_{\rm d}) + r_{1}v_{1}n_{1}\cdot d_{\rm D}(N-n_{1}-n_{2})/(N_{r}N_{\rm d}),\nonumber\\
P^{0,+}(n_{1},n_{2}) &=& r_{2}\omega_{2}n_{2}\cdot d_{\rm D}(N-n_{1}-n_{2})/(N_{r}N_{\rm d}),\nonumber\\
P^{0,-}(n_{1},n_{2}) &=&  \left
( r_{1}u_{1}n_{1} + r_{2}u_{2}n_{2}\right )\cdot d_{2}n_{2}/(N_{r}N_{\rm d}) + r_{2}v_{2}n_{2}\cdot d_{\rm D}(N-n_{1}-n_{2})/(N_{r}N_{\rm d}),\nonumber\\
P^{+,+}(n_{1},n_{2}) &=& 0, \nonumber\\
P^{-,-}(n_{1},n_{2}) &=& (v_{1}r_{1}d_{2} + v_{2}r_{2}d_{1})n_{1}n_{2}/(N_{r}N_{\rm d}),\nonumber\\
P^{+,-}(n_{1},n_{2}) &=& \omega_{1}r_{1}n_{1}\cdot d_{2}n_{2}/(N_{r}N_{\rm d}),\nonumber\\
P^{-,+}(n_{1},n_{2}) &=&  \omega_{2}r_{2}n_{2}\cdot d_{1}n_{1}/(N_{r}N_{\rm d}),\nonumber\\
P^{-2,0}(n_{1},n_{2}) &=& v_{1}r_{1}d_{1}n^{2}_{1}/(N_{r}N_{\rm d}),\nonumber\\
P^{0,-2}(n_{1},n_{2}) &=& v_{2}r_{2}d_{2}n^{2}_{2}/(N_{r}N_{\rm d}),
\label{transition}
\eea

\nd where $P^{-2,0}$ is the probability to gain two $S_{1}$ in a generation. Similarly, $P^{0,-2}$ stands for the probability to gain two $S_{2}$'s. Notice that due to constant population assumption we have replaced differentiated cell population $n_{\rm D}$ by $N - n_{1} - n_{2}$.

The probability of composite events from Fig. \ref{asymmetric} modifies only the transition probability, $P^{0,+}$,

\be
P^{0,+} \rightarrow P^{0,+} + q(N- n_{1}-n_{2})^{2}/(N_{r}N_{\rm d}),
\ee
\nd while the denominator, $N_{\rm r}$, is changed to,

\be
N_{r} = r_{1}n_{1} + r_{2}n_{2} + q(N-n_{1}-n_{2}),
\ee

\nd applied to all $P(\pm,\pm)$'s.

The dynamics of average values for frequencies $\langle x_{1}(t) \rangle$ and $\langle x_{2}(t) \rangle$ can be derived using the master equation,
by inserting average frequencies as,

\bea
\langle x_{1}(t) \rangle = \frac{1}{N} \sum_{n_{1},n_{2}} n_{1}\phi(n_{1}, n_{2}; t),\nonumber\\
\langle x_{2}(t) \rangle = \frac{1}{N} \sum_{n_{1},n_{2}} n_{2}\phi(n_{1}, n_{2}; t),\nonumber\\
\eea

\nd into master equation,

\bea
&&\frac{1}{N}\frac{\partial}{\partial t}\sum_{n_{1},n_{2}}n_{1}\phi(n_{1},n_{2};t)\nonumber\\
&=&\sum_{n_{1},n_{2}}n_{1}( \sum_{\alpha,\beta}P^{\alpha,\beta}(n_{1}-\alpha, n_{2}-\beta)\phi(n_{1}-\alpha, n_{2}-\beta;t))\nonumber\\ &-&(\sum_{\alpha,\beta}P^{\alpha,\beta}(n_{1},n_{2}))\phi(n_{1}, n_{2};t),
\eea

\nd where $\alpha, \beta = \{ \pm 1, \pm 2 \}$ and $\phi(-1,n_{2}) = \phi(n_{1},-) = \phi(-2, n_{2}) = \phi(n_{1},-2) = 0$. By shifting the sum over
$n$'s by $\alpha$ and $\beta$ in the first term of above equation we arrive at ,

\bea
\frac{{\rm d} \langle x_{1}(t)\rangle}{{\rm d} t} &=& \langle a_{1}(x_{1},x_{2})\rangle,\nonumber\\
\frac{{\rm d} \langle x_{2}(t)\rangle}{{\rm d} t} &=& \langle a_{2}(x_{1},x_{2}) \rangle, \nonumber\\
\eea

\nd where,

\bea
a_{1}(x_{1}, x_{2}) &=& \sum_{\alpha}P^{+, \alpha} - P^{-,\alpha} - 2P^{-2,0}, ~~~~ \alpha = \left\{ 0, + , - \right\}, \nonumber\\
a_{2}(x_{1}, x_{2}) &=& \sum_{\alpha}P^{\alpha, +} - P^{\alpha, -} -  2P^{0,-2}, ~~~~ \alpha = \left\{ 0, + , -\right \}.\nonumber\\
\eea

This will lead to a much simpler form for the time evolution of average population frequencies. Thus in the large population limits we have Eq. \ref{replicator}.

\section{Appendix B: Analytical Result for Time to Fixation}

Eq. \ref{replicator} can be exactly integrated to give the time to fixation. For $d_{1,2,\rm D} = 1.0$ we obtain,

\be
\displaystyle
\ln \left(\frac{x_{1,f}}{x_{1,i}} \right) - \frac{1}{r}\ln \left(\frac{x_{2,f}}{x_{2,i}} \right) = \frac{(r-1)\cdot t}{r}
\ee

\nd For a fixation process values of initial and final conditions are,

\bea
\begin{array}{cc}
\displaystyle x_{2,i} = \frac{1}{N} & \displaystyle x_{1,i} = 1 - \frac{1}{N}  \\
\displaystyle x_{2,f} = \alpha \left(1 - \frac{1}{N}\right)  & \displaystyle x_{1,f} = \frac{\alpha}{N}
\end{array}
\label{init}
\eea

\nd These conditions come from the constraints $x_{1,i} + x_{2,i} + x_{{\rm D},f} = 1$. The initial number of differentiated cells are assumed to be zero, and the final number of differentiated cells is assumed to be $x^{*}_{\rm D} = 1- \alpha$ while assuming $x_{1,f} = \alpha/N$.

\nd The fixation time now reads,

\bea
t_{\rm F} &=& \frac{r+1}{r -1}\cdot \ln N -\ln \alpha,\nonumber\\
t_{\rm F} &=& t_{\rm Moran} -\ln \alpha.
\eea

Notice that the the second term which represents the effect of differentiation, is positive as $\alpha = 1 - u - 2v$ is always less than one. We also have $t_{\rm F} \rightarrow \infty$ as $\alpha \rightarrow 0$, which is the case when $u,v \rightarrow 1/3$.

Similar result can be obtained for the non-symmetric situation where the differentiation rates are not equal, i.e., $\alpha_{1} \neq \alpha_{2}$. Integrating the above equation and using the initial conditions similar to Eq. \ref{init},

\bea
\begin{array}{cc}
\displaystyle x_{2,i} = \frac{1}{N} & \displaystyle x_{1,i} = 1 - \frac{1}{N}  \\
x_{2,f} = \alpha_{2} \left(1 - \frac{1}{N} \right)  & \displaystyle x{1,f} = \frac{\alpha_{2}}{N}
\end{array}
\label{init2}
\eea

\nd The (approximate) result for the fixation time is,

\be
t_{F} = \left(\frac{r\alpha_{2} + \alpha_{1}}{r\alpha_{2}-\alpha_{1}}\right)\ln N - \ln \alpha_{2}.
\ee


\bibliographystyle{spbasic}      
\bibliography{mybib-CSC.bib}   

\end{document}